\pdfoutput=1
\documentclass{emulateapj}
\usepackage{multirow}

\newbox\grsign \setbox\grsign=\hbox{$>$}
\newdimen\grdimen \grdimen=\ht\grsign
\newbox\laxbox \newbox\gaxbox
\setbox\gaxbox=\hbox{\raise.5ex\hbox{$>$}\llap
     {\lower.5ex\hbox{$\sim$}}}\ht1=\grdimen\dp1=0pt
\setbox\laxbox=\hbox{\raise.5ex\hbox{$<$}\llap
     {\lower.5ex\hbox{$\sim$}}}\ht2=\grdimen\dp2=0pt

\newcommand{\eqon}{\begin{equation}}
\newcommand{\eqoff}{\end{equation}}
\usepackage[usenames]{color}

\newcommand{\black}[0]{\color{black}}
\newcommand{\red}[0]{\color{black}}

\usepackage{amsmath}
\shorttitle{Relativistic jet formation}
\shortauthors{Porth \& Fendt}

\begin{document}

\title{Acceleration and collimation of relativistic MHD disk winds}

\author{Oliver Porth\altaffilmark{1}}
\author{Christian Fendt 
       }
\affil{ Max Planck Institute for Astronomy, K\"onigstuhl 17,
D-69117 Heidelberg, Germany}
\email{porth@mpia.de, fendt@mpia.de}

\altaffiltext{1}{Fellow of the International Max Planck Research School for
 Astronomy and Cosmic Physics at the University of Heidelberg (IMPRS-HD)}

\begin{abstract}
We perform axisymmetric relativistic magnetohydrodynamic (MHD) simulations
to investigate the acceleration and collimation of jets and outflows from
disks around compact objects.
Newtonian gravity is added to the relativistic treatment in order to
establish the physical boundary condition of an underlying accretion disk
in centrifugal and pressure equilibrium. The fiducial disk surface 
(respectively a slow disk wind) is prescribed as boundary condition for 
the outflow. We apply this technique for the first time in the context of relativistic jets.  The strength of this approach is
that it allows us to run a parameter study in order to investigate how the
accretion disk conditions govern the outflow formation.  
Substantial effort has been made to implement a current-free,
numerical outflow boundary condition in order to avoid artificial collimation
present in the standard outflow conditions.
Our simulations using the PLUTO code run for 500 inner disk rotations and
on a physical grid size of 100x200 inner disk radii.
The simulations evolve from an initial state in hydrostatic equilibrium and
an initially force-free magnetic field configuration.
Two options for the initial field geometries are applied - a hourglass-shaped
potential magnetic field and a split monopole field.
Most of our parameter runs evolve into a steady state solution which can
be further analyzed concerning the physical mechanism at work.
In general, we obtain collimated beams of mildly relativistic speed with
Lorentz factors up to 6 and mass-weighted half-opening angles of 3-7 degrees.
The split-monopole initial setup usually results in less collimated
outflows.
The light surface of the outflow magnetosphere tends to align 
vertically - implying three relativistically distinct regimes in the 
flow - an inner sub-relativistic domain close to the jet axis, 
a (rather narrow) relativistic jet and a surrounding subrelativistic 
outflow launched from the outer disk surface - similar to the spine-sheath
structure currently discussed for asymptotic jet propagation and stability.
The outer subrelativistic disk-wind is a promising candidate for the X-ray absorption winds that are observed in many radio-quiet AGN.
The hot winds under investigation acquire only low Lorentz factors due to the rather high plasma-$\beta$ we have applied in order to provide an initial force-balance in the disk-corona.  
When we increase the outflow Poynting flux by injecting an additional 
disk toroidal field into the outflow, the jet velocities achieved are higher.  These flow gains super-magnetosonic speed
and remains Poynting flux dominated.
\end{abstract}


\section{Introduction}
Astrophysical jets emanate from sources spanning a huge range in energy output 
or length scale - among them young stellar objects (YSO), 
stellar mass compact objects as X-ray binaries or $\mu$-quasars, 
or the powerhouses of some active galactic nuclei (AGN) which host a super-massive
black hole.
In particular for radio-loud quasars, 
for which synchrotron emission dominates the radio spectrum,
relativistic jets are a generic feature. 
Due to the omnipresent angular momentum conservation, mass accretion to all of
these objects features a disk structure around the central mass.
It is commonly believed that jets are launched as disk winds, 
which are further accelerated and collimated by magnetic forces
(see \citet{1982MNRAS.199..883B, 1983ApJ...274..677P, 1986A&A...162...32C, 1997SvPhU..40..659B, 
2003ApJ...596.1240H, 2007prpl.conf..277P}.
Relativistic jets may gain further energy by interaction with the
black hole magnetosphere 
\citep{1977MNRAS.179..433B, 1997MNRAS.292..887G, 2005MNRAS.359..801K}.

The magnetohydrodynamic (MHD) self-collimation of non-relativistic jets has been 
proven in general by 
time-dependent simulations \citep{1995ApJ...439L..39U,1997ApJ...482..712O} and 
have been investigated in further detail considering additional physical effects 
as magnetic diffusivity by 
\cite{2002A&A...395.1045F}, a variation in \cite{1999MNRAS.309..233O}, 
non-axisymmetric instabilities in the launching region \citep{2003ApJ...582..292O},
or a variation in the mass flow profile or the magnetic field geometries
\citep{2006ApJ...651..272F, 2006MNRAS.365.1131P},
or the influence of a central magnetic field \citep{2009ApJ...692..346F,2008A&A...477..521M}.

In the case of relativistic jets the efficiency of MHD self-collimation is under debate. 
The main reason is the existence of electric fields which are negligible for non-relativistic
MHD and which are commonly thought to have a net de-collimating effect on the jet.
Essentially, \cite{1991ApJ...377..462C} have demonstrated the current carrying 
relativistic jet can be highly collimated.
However, the actual structure of these jets still remains unclear - mainly due to 
the need for simplifying assumptions to solve the corresponding set of MHD equations.

So far, a variety of theoretical models have been developed for the case of 
{\em self-similar} jets
\citep{1992ApJ...394..459L, 1994ApJ...432..508C, 1995ApJ...446...67C,
      2003ApJ...596.1104V,2006A&A...447..797M},
although it seems clear that relativity does not obey self-similarity.  
Fully 2.5D theoretical solutions for the internal magnetic jet structure could be
obtained by neglecting matter inertia
\citep{1997A&A...319.1025F, 2001A&A...365..631F}.
These force-free solutions for the field structure can in principle be coupled
to the dynamical wind solution along the field lines
\citep{1996A&A...313..591F, 2001A&A...369..308F, 2004ApJ...608..378F}.
\cite{1997A&A...319.1025F} obtained solutions for the internal jet force balance 
in Kerr metric with an asymptotically cylindrical jet emerging from a disk-like
structure around the central rotating black hole. 
The shape of the collimating jet boundary was obtained as result of the internal 
force equilibrium, in particular considering the regularity condition along the 
jet outer light surface.

Time-dependent simulations of relativistic MHD jet formation have been performed
considering a general relativistic metric, including also the evolution of 
the underlying accretion disk.
Early - seminal - simulations did last for a few inner disk rotations 
only \citep{1998ApJ...495L..63K, 1999ApJ...522..727K}, which is
sufficient time to demonstrate the launching of an outflow, 
but hardly sufficient in order to investigate the long-term dynamical evolution of
the emerging jet.

More recent simulations were able to follow several 100 disk rotations and show 
the formation of a so-called funnel flow origination in the shear layer between 
the horizon and the inner disk radius
\citep{2005ApJ...620..878D, 2007MNRAS.375..513M, 2008MNRAS.388..551T, 2009MNRAS.394L.126M}.
These simulations indicate a highly time-variable mass ejections of rather low degree 
of collimation. However, the funnel flow achieves Lorentz factors of up to 50.
General relativistic MHD simulations are also able to determine the interrelation 
between jet formation and the Blandford-Znajek mechanism 
\citep{2005ApJ...630L...5M, 2007MNRAS.377L..49K}.

Ultra-relativistic MHD simulations of accelerating and collimating
jets have been presented by \citet{2007MNRAS.380...51K},
spanning over a huge range of length scale and providing jets of 
large Lorentz factor $\Gamma \sim 10$.
Their simulations, however, did not start from the very base of the 
jet - the accretion disk, but at some fiducial boundary above the equatorial
plane.
Since the jet has been launched already with super-escape speed,
gravity has not been considered. 
The jet flow has been confined within a rigid wall of predefined shape 
which naturally affects the opening angle of the MHD jet nozzle and thus
jet collimation and acceleration.

The focus of our present paper is 
i) to concentrate on the formation and acceleration of a relativistic MHD jet right from
the launching area the accretion disk surface,
ii) to investigate the (self-) collimation of relativistic MHD jets under the influence
of de-collimating electric forces and an "open" boundary condition for the outflow 
iii) to consider gravity as en essential gradient to provide a realistic disk boundary condition in
equilibrium,
iv) to run long-term simulations lasting more than 1000 inner disk rotations until the
jet reaches steady state,
v) to concentrate on MHD disk jets as disks are the natural origin for the mass load for
AGN jets.

The outline of this paper is as follows: In section \ref{sec:concepts} we discuss the 
concepts of ideal special relativistic magnetohydrodynamics in the
perspective of jet formation.  
Section \ref{sec:model setup} is devoted to the initial- and boundary conditions of the
numerical simulations, whose results are shown in section \ref{sec:results}. 
We conclude in section \ref{sec:conclusions}.

\section{Concepts of relativistic MHD jets}\label{sec:concepts}

It is well known that relativistic jets must be strongly magnetized 
\citep{1969ApJ...158..727M, 1986A&A...162...32C, 1993ApJ...415..118L}.
This simply reflects the fact that the lower the mass flux, the more electro-magnetic energy (Poynting flux) 
can be transferred into high kinetic energy per unit mass.
Relativistic MHD is also limited for very strong magnetization as then
the MHD assumption can be violated since a sufficiently large amount of electric charges is lacking which are needed to drive the electric current system. Such a situation might arrise in the ultra-relativistic regime of pulsar-winds but is unlikely for the disk-winds investigated here.  

This paper deals with the time-dependent formation of relativistic MHD jets 
by using the special relativistic MHD module of the PLUTO code provided by \cite{2007ApJS..170..228M}
and applying a Newtonian description of gravity. 
%
\subsection{Relativistic MHD equations}

The relativistic MHD module of PLUTO solves the system of special relativistic conservation laws.
In a covariant formulation, 
the equations follow naturally as a set of hyperbolic equations.
It is solved for energy and momentum conservation
\eqon
\partial_{\alpha} T^{\alpha \beta} = 0 \label{eq:energy}
\eqoff
of an ideal magnetized fluid 
\eqon
T^{\alpha \beta} = (\rho h + b^2) u^\alpha u^\beta + 
\left(p + \frac{1}{2} b^2\right) g^{\alpha \beta} -b^\alpha b^\beta
\eqoff
with the specific plasma enthalpy 
\eqon
h=\frac{\gamma}{\gamma-1}\frac{p}{\rho}+1,
\eqoff
the isotropic gas pressure $p$, density $\rho$ (both in the local rest-frame), four-velocity $(u^\alpha)=(\Gamma,\Gamma \mathbf{\beta})^T$, velocity $\mathbf{\beta}=\mathbf{v}/c$, Lorentz factor $\Gamma=(1-\beta^2)^{-0.5}$ and the magnetic field pseudo vector
\eqon
b^\alpha = -\frac{1}{2} \epsilon^{\alpha \beta \gamma \delta} u_{\beta} F_{\gamma \delta}.
\eqoff

Assuming infinite conductivity, thus vanishing electric fields in the 
rest-frame of the plasma $F^{\alpha \beta} u_{\beta}= 0$, the 
homogenous Maxwell equation
\eqon
 \partial_{\alpha}\ ^{*}\!F^{\alpha \beta} = 0 \label{eq:hommaxwell}
\eqoff
can be written solely in terms of the magnetic four-vector $b^\alpha$
\eqon
^{*}\!F^{\alpha \beta} \equiv \frac{1}{2} \epsilon^{\alpha \beta \gamma \delta} F_{\gamma \delta} 
                     =   b^\alpha u^\beta - b^\beta u^\alpha\label{eq:hommhd}
\eqoff
and for the field vector components it 
follows\footnote{following the convention that Greek indices run 
from 0 to 4 whereas Latin indices go from 1 to 3}
\begin{eqnarray}
B^{i} &=& ^{*}\!F^{i 0} = b^{i} u^0 - b^{0}u^{i}\\
E^{i} &=& \epsilon^{ijk}b^j u^k \label{eq:mhd-cond}.
\end{eqnarray}
Equation \ref{eq:mhd-cond} represents the ideal MHD condition 
$\mathbf{E} = - \mathbf{\beta} \times \mathbf{B}$ 
and is the reason why all electric fields can be eliminated from the equations.  
The magnetic four-vector turns out as $b^0 = B^{i} u^{i};\ \  b^{i} = (B^{i}+b^0 u^{i})/u^0$.  
The conservation of the Faraday tensor given by Eq.\,\ref{eq:hommhd} results in the 
non-relativistic (ideal) induction equation
and the solenoidal condition $\mathbf{\nabla} \cdot \mathbf{B} = 0$.  
Mass conservation is guaranteed by the continuity equation 
\eqon
\partial_{\alpha} (\rho u^\alpha) = 0. \label{eq:continuity}
\eqoff
We apply a polytropic equation of state for the gas with the polytropic
index $\gamma =5/3$.

\subsection{Gravity in special relativity}
The outcome of MHD simulations is mainly determined by its boundary conditions and therefore requires great care in describing the proper physical state of 
interest.  
Since in our simulations the jet is considered to be launched as a wind from a 
rotating disk,
it is essential to take into account a proper disk model as boundary condition.
For the disk boundary we choose a (sub-) Keplerian rotation profile and a
hydrostatic pressure gradient (see section \ref{sec:injection}).
This choice implies a hot, ``puffed up'' corona with a sound speed 
$c_{s}^2\simeq 2/3\ v_{\phi}^2$.
This is the main reason why we have added gravity to our special relativistic
treatment.
The direct impact of gravity on the jet dynamics is marginal as the outflow
is accelerated to super escape speed quickly.

A fully self-consistent relativistic treatment of gravity would imply
a general relativistic approach.
However, we are not interested in the region very close to the horizon 
of the central object, but in the long-term dynamics and evolution of a disk wind into
a relativistic jet.
We can therefore neglect general relativistic effects in our simulation 
domain.

We apply a softened gravitational potential
\eqon
\phi = -\frac{GM}{R+r_{S}}
\eqoff
with a softening length of $r_{S}=1/3$ that may be related to the to the Schwarzschild 
radius of a non-rotating black hole\footnote{The capital $R=\sqrt{r^2+z^2}$ denotes the spherical radius throughout this work.}.
\red
The corresponding acceleration reads
\eqon
\mathbf{a} = -\nabla\phi = -\frac{GM}{(R+r_{S})^2 R}\mathbf{r}
\eqoff
and hence instead of solving equation \ref{eq:energy}, we solve 
\eqon
\partial_{\alpha}T^{\alpha \beta} = f^\beta
\eqoff
with the four force density 
$
	(f^\beta) = \Gamma \rho (\mathbf{a \cdot v},\mathbf{a})^T
$
as a local source term on the right-hand side.  
This is incorporated in PLUTO as a "body force" ($\Gamma \mathbf{a}$) using the infrastructure of the code.  
\black

Omission of softening would lead to numerical errors 
(due to the unresolved steep gradients in the potential close to the origin), 
piling up to produce artificial acceleration along the spine of the jet close to 
the axis.
Softening is clearly a compromise avoiding the singularity (by limiting the 
required resolution) on little cost of realism.  
Another choice could be the well-known pseudo-potential by \cite{1980A&A....88...23P}
which has just the negative softening $\phi_{PW}=-GM/(R-r_{S})$.  
For the cylindrical geometry of our choice, the singularity would become even more problematic,
complicating the setup a great deal.

\subsection{Relations in axisymmetric MHD}
The region of jet formation may be fairly well approximated in {\em axisymmetry}.
In fact, non-axisymmetric distortions may actually hinder the formation of powerful jets as probably 
demonstrated by the existence of a variety of strongly magnetized, rapidly rotating accretion disk 
systems which, however, do not exhibit jets (e.g. cataclysmic variables or most pulsars). 

Under the assumed symmetry in a cylindrical coordinate system, 
the magnetic field vector can be written as
\begin{equation}
	\mathbf{B} = \mathbf{B}_{p} + B_{\phi} \mathbf{e_{\phi}},
\end{equation}
where $B_{\phi}$ can now be an arbitrary function of $r$ and $z$, as the solenoidal condition
translates to $\mathbf{\nabla\cdot B_{p}}=0$. 
The stream function 
$\Psi(r,z)= (1/2\pi) \int \mathbf{dS \cdot B_{p}} = rA_{\phi}$ 
measures the magnetic flux through the surface area $S$ 
and follows from the toroidal component of the vector potential,
\begin{equation}
\mathbf{B_{p}}= \mathbf{\nabla}\times \mathbf{A_{\phi}} 
              = \mathbf{\nabla}\times \frac{\Psi\ \mathbf{e_{\phi}}}{r} 
              = \frac{1}{r}\mathbf{\nabla}\Psi \times\ \mathbf{e_{\phi}}
\end{equation}

For the electric field, the ideal MHD condition $\mathbf{E = B \times \beta}$ gives 
\begin{equation}
\mathbf{E} = \frac{r\Omega^{F}}{c} B_{p} \mathbf{n} = \frac{r}{r_{\rm L}} B_{p}\mathbf{n}\label{eq:electric-field}
\end{equation}
in terms of the so called angular velocity of the field line 
$\Omega^F = \left(v_{\phi} - v_{p}B_{\phi}/B_{p}\right)/r$, 
or the so-called light cylinder radius \footnote{This is the radius at which the hypothetical angular 
velocity of a field line supercedes the speed of light} of a field line $r_{\rm L} \equiv c/\Omega^F$.  
The direction of the electric field is given by 
$\mathbf{n = B_{p}}/B_{p}\times \mathbf{e_{\phi}}$ and is perpendicular to the magnetic
flux surface $\Psi(r,z)$.  
The poloidal Poynting flux $\mathbf{\mathcal{S}} = (c/4\pi)\mathbf{E\times B_{\phi}}$ 
simplifies to 
\begin{equation}
	\mathbf{\mathcal{S}} = -\frac{c}{4\pi} \frac{r}{r_{\rm L}} B_{\phi}\mathbf{B_{p}} 
                             = -r\Omega^{F} \frac{B_{\phi}\mathbf{B_{p}}}{4\pi}.
\end{equation}

\subsubsection{Perpendicular and parallel force-balance}\label{sec:force-balance}
\begin{figure}[htbp]
\centering
\includegraphics[width=6cm]{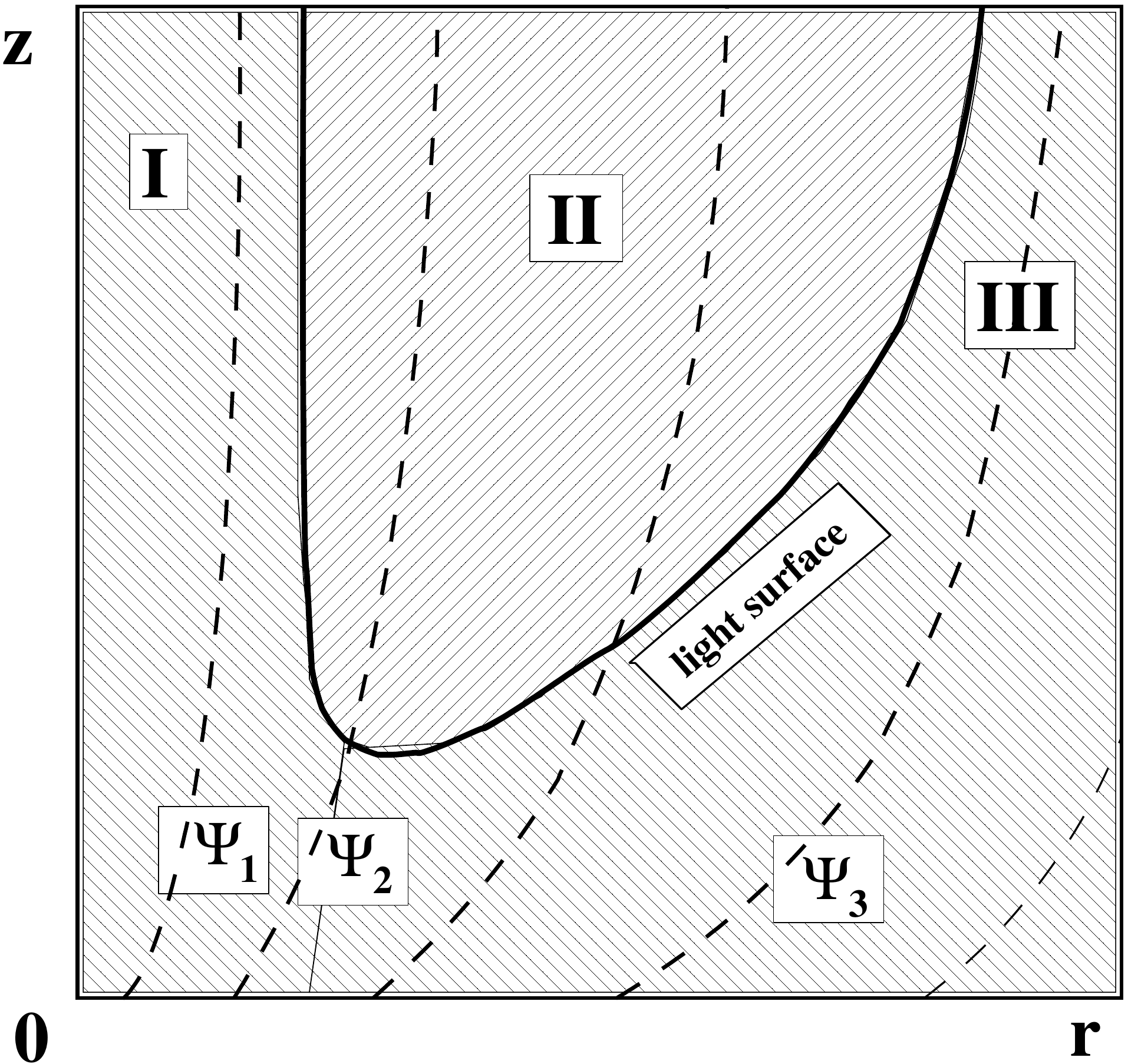}
\caption{The different dynamical regimes in special relativistic
disk-winds. Region I and III stay sub-relativistic, while region II
is relativistic, i.e. electric forces are not negligible
(see Sect.\,\ref{sec:force-balance} for a discussion).
\label{fig:lightcylinder}}
\end{figure}

The processes leading to flow collimation can be identified directly 
from the (steady-state) trans-field force-balance equation 
\citep{1991ApJ...377..462C, 1993A&A...270...71A}.
Here, we adopt the notation of the latter paper when investigating the 
collimation behavior in the quasi steady-state time domain of our 
simulations.
The curvature 
$\kappa \equiv \mathbf{n}\left( \mathbf{B_{p}} \cdot \nabla \right) \mathbf{B_{p}} /B_{p}^2$
of a flux surface $\Psi(r,z)$ results from the summation of perpendicular forces,
\begin{equation}
\begin{split}
\kappa \frac{B_{p}^2}{4\pi}\left(1-M^2-\frac{r^2\Omega^{F^2}}{c^2}\right) =  \\
+ \left(1-\frac{r^2\Omega^{F^2}}{c^2}\right)\nabla_{\bot}\frac{B_{p}^2}{8\pi} 
+ \nabla_{\bot}\frac{B_{\phi}^2}{8\pi}+\nabla_{\bot}p \\
+ \left(\frac{B_{\phi}^2}{4\pi r}-\frac{\rho h u_{\phi}^2}{r}\right)\nabla_{\bot}r
- \frac{B_{p}^2\Omega^{F}}{4\pi c^2}\nabla_{\bot}\left(r^2\Omega^F\right) \\
+ \Gamma \rho \nabla_{\bot}\phi, 
\end{split}
\label{eq:trans-field}
\end{equation}
where we have added the collimating component of the gravitational force. 
For the ease of use in section \ref{sec:trans-field}, we label the terms as 
$(F_{\rm curv}, F_{\rm pbp}, F_{\rm pbphi}, F_{\rm p}, F_{\rm pinch}, 
F_{\rm cf}, F_{\rm el}, F_{\rm grav})$ 
in the order of their appearance in Eq.\ref{eq:trans-field}.
The (poloidal) Alfv\'en Mach number $M$ is relativistically defined as 
\begin{equation}
	M^2 = \frac{4\pi \rho h u_{p}^2}{B_{p}^2}.
\end{equation}
The gradient $\nabla_{\perp}\equiv\mathbf{n}\cdot \nabla$ 
is projected perpendicular to the magnetic flux surfaces $\Psi$, and thus along the 
(inward pointing) electric field.  
The light surface of a magnetosphere is located where 
$r_{\rm L}(\Psi) = r_{\rm L}(r,z) \equiv c / \Omega^F(\Psi)$
\footnote{Thus, the light surface consists of the points 
of intersection between the field lines with their corresponding light cylinder}, it hence depends on the flux-geometry as well as on the rotation profile $\Omega^F$.  
%
Each flux surface / magnetic field line crosses the light surface at most once (see
also the discussion in \citet{1997A&A...323..999F})
Some field lines $\Psi(r,z)$ never cross the light surface, indicating an asymptotic radius
$r_{\infty}(\Psi) < r_{\rm L,\Psi}$. For these field lines relativistic effects due
to rotation (electric fields) are less important.
For others, the asymptotic radius is $r_{\infty}(\Psi) > r_{\rm L, \Psi}$.
The light surface constitutes a critical point of the stationary axisymmetric wind equation only
in the mass-less limit in which it is identical to the modified poloidal Alfv\'en surface 
$M_{A}^2 = 1-(r_{A}/r_{L})^2$ \citep{1986A&A...156..137C, 1986A&A...162...32C}. 
However, it is essential to note that at the light surface the dynamical behavior of 
the poloidal magnetic pressure term changes - the force changes sign.
This leads to the existence of three dynamically different regimes in the asymptotic
(collimated) region of a relativistic jet (see Fig.\,\ref{fig:lightcylinder}).
In region I, for all field lines $r_{\infty}(\Psi) < r_{\rm L, \Psi}$ corresponding
to $(1-(r/r_{\rm L})^2)>0$, and, thus, a de-collimating magnetic pressure term.
Field lines in region II do cross their light cylinder, and, since $r> r_{\rm L}$,
the magnetic pressure term acts as collimating for $r > r_{\rm L}$.
Field lines in Region III  never reach their light cylinder,
and here the magnetic pressure term is de-collimating again.
The slope of the outer part of the light surface critically depends on the dynamics
and the magnetic field structure of the outflow in the very inner part.

Similarly, the forces due to the electric field $E = r/r_{\rm L} B_{\rm p}$ 
(second last term in
Eq.\,\ref{eq:trans-field}) scale with the relative position to the light surface,
hence they are important in region II of the jet formation region only.

Equation \ref{eq:trans-field} together with Fig.\,\ref{fig:lightcylinder} once
more demonstrates the need to resolve the whole acceleration and collimation
region of a jet in radial and vertical direction. 
Only when the light surface is taken into account self-consistently, the proper
force-balance is applied along and across the flow. 

Similarly one can derive the parallel-field force equation, it becomes:
\eqon
\begin{split}
	\frac{B_{p}^2}{4\pi}\nabla_{||}M^2 = 
	\kappa_{||} \frac{B_{p}^2}{4\pi}\left(1-M^2\right) \\
	-\nabla_{||}\left(p+\frac{B_{p}^2}{8\pi}+\frac{B_{\phi}^2}{8\pi}\right) \\
	- \left(\frac{B_{\phi}^2}{4\pi r}-\frac{\rho h u_{\phi}^2}{r}\right)\nabla_{||}r \\
	- \Gamma \rho \nabla_{||} \phi
	\label{eq:parallel-field}
\end{split}
\eqoff
with the necessary definitions $\nabla_{||}\equiv\mathbf{B_{p}}/B_{p}\nabla$ and 
$\kappa_{||}B_{p}^2\equiv 
\mathbf{B_{p}}/B_{p}
\left(
\mathbf{B_{p}}
\cdot 
\nabla
\right)
\mathbf{B_{p}}
$.
We see how the change in the Mach-number is mediated by the interplay of tension-, pressure-,
pinch-, centrifugal- and gravitational- acceleration.   Electric fields (pointing in the
perpendicular direction) don't contribute and the equation reduces to the Newtonian case:
In steady-state, there is no electric acceleration!

\red
A number of self-similar approaches to the relativistic
 jet formation have beed published (eg. \cite{2003ApJ...596.1104V}).
 While the self-similar ansatz is a powerfull and highly
 successfull tool to solve the non-relativistic MHD problem
 (starting with the Blandford-Payne solution), we believe
 that using self-similarity for relativistic MHD jets is
 problematic.
 
We note that neither the light surface nor the relativistic Alfv\'en
surface obeys a self-similar structure.
It is well known that forcing self-similarity into the relativistic 
MHD equations constrains
the rotation law for the magnetosphere $\Omega^F(r)  \propto r^{-1}$.
(see also discussion in \citet{1992ApJ...394..459L,1993ApJ...415..118L}).
This is a major difference to the non-relativistic self-similar
approach.
We further note that also the scaling for the electric
field depends on the radial position of the light surface 
(see Eq.~\ref{eq:electric-field}).
This is, however, of uttermost importance for the structure of relativistic
magnetospheres as the electric field forces play a leading role in the
trans-field force-balance (equation \ref{eq:trans-field}).
Similar arguments hold for the inner light surfaces around Kerr 
black holes or the geometry of the black hole ergosphere. 

We therefore believe that a steady-state self-similar relativistic MHD 
approach is intrisically inconsistent with the relativistic 
characteristic of the flow.
\black
\subsubsection{field line constants}\label{sec:flconst}
Stationary axisymmetric MHD flows conserve the following five quantities along
the magnetic flux-function $\Psi$.
From the iso-rotation law together with the ideal MHD condition follows
the rest mass energy-flux per magnetic induction, 
\eqon
k = k(\Psi) \equiv \frac{\rho u_{p}}{B_{p}}\label{eq:kf}
\eqoff
and the iso-rotation parameter 
\eqon
\Omega^F = \Omega^F(\Psi) = \frac{1}{r}\left(v_{\phi}-v_{p}\frac{B_{\phi}}{B_{p}}\right)\label{eq:omegaf}
\eqoff
(often interpreted as angular velocity of the field lines).  
In absence of shocks the (pseudo-) entropy 
\eqon
Q=\frac{p}{\rho^\gamma} = Q(\Psi) \label{eq:Qf}
\eqoff
is conserved as well as the angular momentum flux
\eqon
l=-\frac{I}{2\pi k c} + r u_{\phi} = l(\Psi)\label{eq:lf}
\eqoff
and the flux ratio of total energy to rest-mass energy,
\eqon
\mu = \frac{\mathcal{S+K+M+T+G}}{\mathcal{M}} = \mu(\Psi) \label{eq:mu}
\eqoff
where we identify the individual terms as
(purely) kinetic energy flux
$\mathcal{K} \equiv (\Gamma-1) \rho\ u_{p}$,
rest-mass energy flux
$\mathcal{M} \equiv \rho\ u_{p}$,
thermal energy flux
$\mathcal{T} \equiv \Gamma \frac{\gamma}{\gamma-1} p\ u_{p}$,
and gravitational energy flux
$\mathcal{G} \equiv \rho\ \phi\ u_{p}$,
respectively.  
The cold, asymptotic limit of (\ref{eq:mu}) is particularly of interest, it reads
\eqon
\mu = \Gamma(\sigma+1)
\eqoff
where $\sigma=\mathcal{S/(K+M)}$ is the customarily defined magnetization parameter
- the ratio of Poynting to kinetic flux. 
This simple relation provides a theoretical maximum for the Lorentz-factor 
$\Gamma^{*}=\mu$, when the entire electromagnetic energy is converted into kinetic 
energy. 

The essential point in the quest for relativistic jets is to find a highly energetic 
disk solution with values of $\mu$ beyond the anticipated Lorentz-factor.  
Previous studies obtaining highly relativistic jets by \cite{2007MNRAS.380...51K} do
not start from a realistic disk solution but inject the jet material with 
an artificial rotation profile, 
a high injection speed $v_{\rm p} > 0.5 c $, and 
a density low enough to obtain a high energy flux $12\leq \mu_{\rm max}\leq 18$.   
In the present study we are aiming to improve on the problems just mentioned 
by applying a physical boundary condition as a Keplerian disk-corona in equilibrium.

\red
\subsection{Accretion disk coronae}\label{sec:coronae}
%
It is our ambition to connect the wind solutions to the ambiance of a realistic accretion disk.  
In our simulations, the flow originates in the high entropy atmosphere called a corona.
Optically thin coronae are an integral part in models of the X-ray features of AGN
\citep[e.g.][]{1993ARA&A..31..717M}
and $\mu$-Quasars \citep[e.g.][]{2002MNRAS.332..856N,2003NewAR..47..491M}.

While Compton cooling can provide the observed spectra, the heating mechanism is not easily found.  
Just as in the case of the sun, the coronal heat cannot directly be transfered from the colder photosphere/accretion disk (according to the second law of thermodynamics) and the nature of vertical energy transport is an active field of research.  
External irradiation of flared disks by the central object (or central disk) is certainly present in a multitude of objects \citep[see][for a review]{2008ChJAS...8..302C} but might not be the primary energy source.  
Among the most promising mechanisms we should highlight magnetic reconnection heating as proposed by \cite{1991ApJ...380L..51H}.  

Between the mid-plane and the coronal point of injection in our simulations, ideal MHD can not provide a realistic picture.  
In order for an accretion disk to {\em work}, a torque of viscous or magnetic origin has to be exerted onto the material. 
Additionally, the flux-freezing constraint of ideal MHD must be relaxed since it would lead to an accumulating magnetic pressure that ultimately stops the accretion process.  
Studies modelling both the accretion motion and the super Alfv\'enic jet based on a stationary self-similar approach (e.g. \cite{1993ApJ...410..218W, 1993A&A...276..625F, 1995ApJ...444..848L}) rely on global-scale magnetic fields and ad-hoc assumptions on the viscosity and (anisotropic) magnetic diffusion.  
In order for these models to be stable against strong magnetic compression on the one side and the magneto rotational instability (MRI; see \cite{1998RvMP...70....1B}) on the other, \cite{1995A&A...295..807F} require equipartition for the thermal and magnetic pressure.
\cite{2000A&A...361.1178C} demonstrated the importance of heating for the mass-loading or the jet. 
Time-dependend numerical simulations of these magnetized accretion ejection structures (MAES) were presented by 
\cite{2002ApJ...581..988C, 2004ApJ...601...90C, 2007A&A...469..811Z} 
and adopt a fixed in time anomalous resistivity profile in order to connect the two dynamical regimes.  

Although the stationarity of the aforementioned simulations is possibly hampered by numerical diffusion and low resolution effects, MAES with global-scale fields are to date the most successful models to create collimated outflows.  

It is now widely believed that the source of viscosity and resistivity in weakly magnetized disks is the turbulence seeded by the MRI.  
Local, stratified shearing box simulations by \cite{2000ApJ...534..398M} suggest a quenching of the MRI in the strongly magnetized coronal region, as the  magnetic scale height exceeds the thermal one.  This is in contrast to the equipartition fields proposed for the MAES.  
Magnetic buoyancy of the large-scale fields eventually created by a turbulent dynamo could then provide the coronal heating.  
Typically, the MRI results in toroidally dominated coronal fields with $B_{\phi}^2 > 10 B_{p}^2$ and opening the field lines towards a topology favorable for wind acceleration remains a challenge.  In the context of the solar corona this process is discussed by \cite{2003ApJ...599.1404W}.  
 
By restraining parts of the accretion disk structure, \cite{2003A&A...398..825V} could achieve outflows in time-dependend simulations of disk-corona structures where the magnetic field is sustained by a mean-field dynamo.  
Analytic models of this turbulent disk-corona-outflow connection are however in still its infancy (see the discussion by \cite{2004ApJ...616..669K} and attempts by \cite{2009ApJ...704L.113B}).

\black
\section{Model setup for the MHD simulations}\label{sec:model setup}
%
\red
With the aforementioned considerations we choose the following model for our investigation.  
A global-scale poloidal field favorable of wind acceleration is adopted.  Whether it is advected by the accretion flow or created by an underlying dynamo is not of our concern.  
The jet base resembles a corona in the sense that it is hot (electron  temperature $\sim10^9 \rm K$), has no mechanism of cooling, is non-turbulent (no viscosity) and highly ionized (infinite conductivity).  We choose a Keplerian rotation profile for the field lines.  
The flow starts with sub-escape velocity and we investigate sub-magnetosonic injection where mass loading is determined by the internal dynamics as well as mass-fluxes imposed by the boundary condition.  
We perform axisymmetric special relativistic MHD simulations of jet formation for a set of different magnetic field geometries and field strengths.  In the following we discuss the numerical realization of our problem.
\black
\subsection{Boundary conditions}

\begin{figure}[htbp]
\centering
\includegraphics[width=7cm]{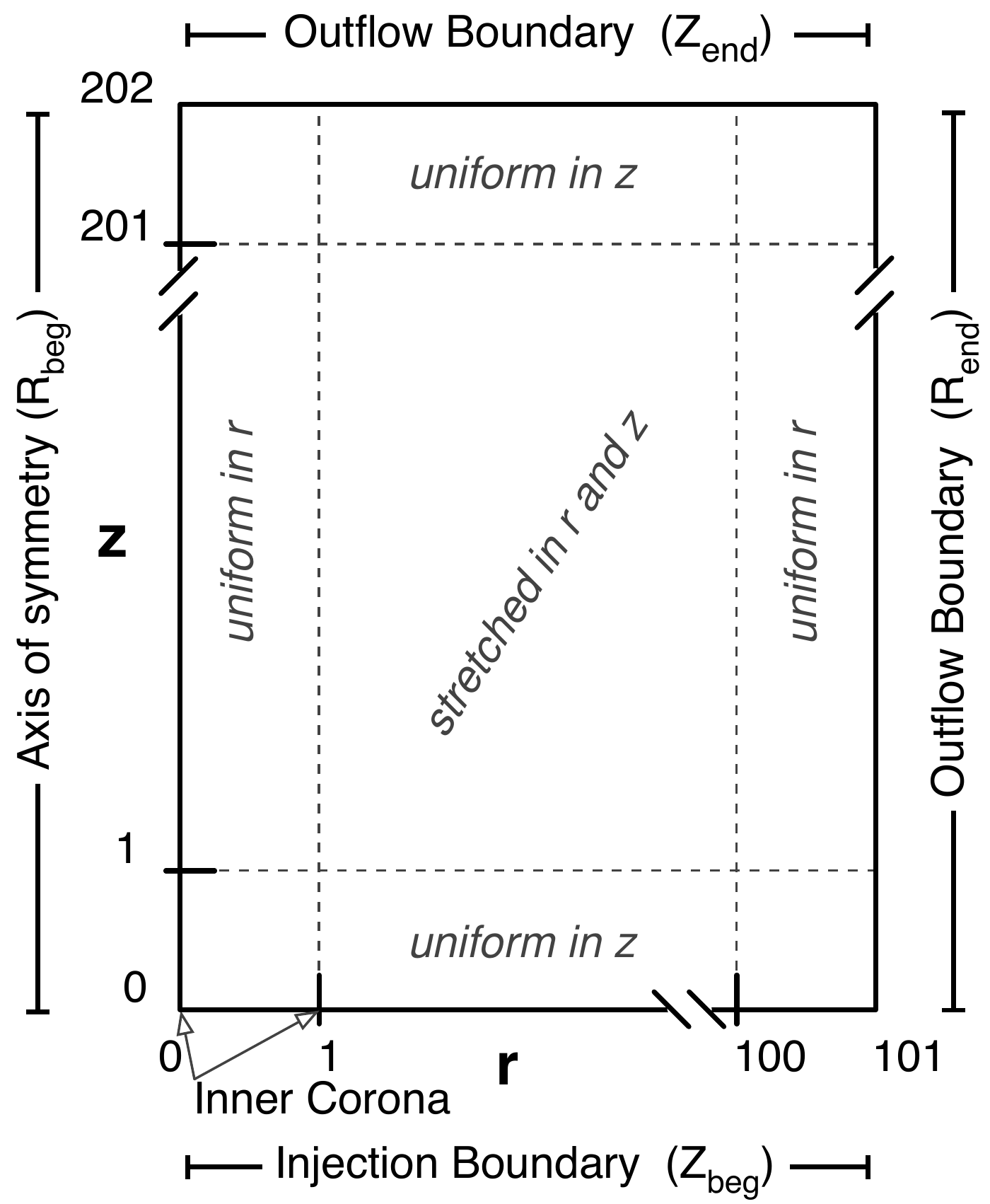}
\caption{Sketch of the different regimes of our grid and boundary.  
In both directions we set 20 equidistant cells in $[0,1]$.
Then follows a stretched grid until we add five equidistant cells one unit-radius before the outflow boundary.  For $r\in[0,1]$, the hydrostatic corona is fixed to minimize the influence of the central region on the disk-wind.  
\label{fig:setup}}
\end{figure}

Given the 2.5 dimensional nature of the problem, three geometrical boundaries have to be 
prescribed. 
These are the inlet boundary along $z=0$ from which material is injected into the domain 
({\em inflow})
and the two outer boundaries at $r=r_{\rm end}$ and $z=z_{\rm end}$ where we expect 
material to leave the computational domain ({\em outflow}). 
The boundary condition along $r=0$ ($\rm R_{beg}$) follows from cylindrical symmetry.  
Figure \ref{fig:setup} gives an overview of the different regions.  

\subsubsection{Injection boundary ($Z_{beg}$)}\label{sec:injection}
Pursuing the aim to follow the acceleration of an disk wind from as close to the accretion 
disk as possible, we start with a sub slow-magnetosonic wind.
We are hence free to choose four constraining boundary conditions without overdetermining
the system (see \cite{Bogovalov1997} and Appendix \ref{sec:MHD-inj} for more details).  

Our choice is to fix the toroidal electric field component $E_{\phi}=0$. 
This suppresses the evolution of the bounding poloidal magnetic field and is realized
by requiring $\mathbf{v_p || B_p}$.  
The field line iso-rotation is kept constant in time and follows a Keplerian rotation law
$\Omega^F\propto r^{-1.5}$.
Unless specified otherwise, the boundary condition starts out initially in a force-free state with zero toroidal field $\Omega^F = v_{\phi}(r)/r$ corresponding to a disk in hydrodynamic equilibrium.  
This is an essential ingredient as - within stationary ideal MHD - $\Omega^{F}$ just 
equals the mid-plane angular velocity of the material $\omega(r)$. 
Relaxation of the infinite conductivity constraint would, however, lead to an inequality 
$\Omega^F(r) \le \omega(r)$ owing to the diffusion of magnetic field. 
For these reasons $\Omega^F$ should closely follow the expected disk rotational profile 
and should be limited by the maximal velocity in the mid-plane, 
typically at the inner edge of the disk located at $r=1$.  

A radial force-equilibrium along the whole boundary is enforced by balancing the centrifugal and 
pressure support against gravity via the sub-Keplerianity of the
rotation $\sqrt{\chi} = v_\phi(r=1)/v_K$ 
\red
that also determines the inlet density, 
\black
where $v_K$ is the circular velocity that alone sustains against gravity at $r=1$.
\red
A more convenient parametrization is in terms of the
	relative temperature 	
	$\epsilon \equiv c_s^2/v_\phi^2 = (\gamma-1)(1-\chi)/\chi$.
\black
We investigate two cases - a hot corona with $\epsilon=2/3$ and a version with $\epsilon=1/6$ 
\red
($\chi=0.5$ and $\chi=0.8$)
\black.

If we interpret $r=1$ as the innermost stable circular orbit (ISCO) around a black hole, $v_{\rm K}$ is a measure of the black hole spin.  In the case of a Schwarzschild black hole it is $v_{K}\simeq0.6c$ while we choose the scaling velocity $v_{\phi}(r=1) = 0.5 c$ for convenience.  

The inner disk-edge is numerically difficult to model because of the transition to the inflow of the disk-wind and the steep gradients in gravity.  Within $r<1$, the so-called plunge region, a physical solution would allow for (radial and vertical) accretion onto the central object. 
Since the dynamics in this area would then require a general relativistic treatment  which we cannot provide in this context, we simply minimize the dynamical effect of this region by freezing the hydrostatic solution initially present in the domain.  
Other authors have assumed a thin funnel flow along the axis (e.g. \cite{1999ApJ...526..631K}) or added an internal sink-cell (like \cite{2002ApJ...581..988C}) to circumvent this problem 
The transition between the ``inner corona'' and disk-wind is smoothed via the Fermi step function 
$F(x) = (1+e^{(1-x)/0.1})^{-1}$.  
Rotational support is thus turned on by the setting 
\begin{eqnarray}
\rho_{\rm d}(r,z) &=& \frac{1}{1-\chi F^2(R)} (R+r_{S})^{1/(1-\gamma)}\label{eq:rhodisk},\\
v_{\phi}(r) &=& \sqrt{\chi} v_{K} F(R) R^{-0.5}.
\end{eqnarray}
Density given by equation \ref{eq:rhodisk} and the coronal pressure $p$ constitute the third and fourth fixed in time conditions.  
%

We emphasize that is is not possible to specify both injection velocity and density-profile and thus the mass-flux for sub-(magneto)sonic flows, 
as this is determined by the sonic point.  
Therefore we match the vertical velocity $v_{z}$ to the domain via $\partial_{z} v_{z} = 0$, 
while the radial component follows from the $E_{\phi}=0$ condition. 
We limit the injection speed by the local slow magnetosonic speed in the case when 
the velocity just above the boundary becomes trans-sonic.
This provides the fifth constraint needed in that case.  

With the induction of a toroidal magnetic field component in the jet, the rotational velocity 
needs to be adjusted in order to satisfy Eq.\,\ref{eq:omegaf},
\eqon
v_{\phi} = r\Omega^F + \frac{v_{p}}{B_{p}}B_{\phi}
\eqoff
as we apply the condition $\partial_{z}B_{\phi} = \partial_{z} B_{r} = 0$ 
(while $B_{z}$ then follows from $\mathbf{\nabla\cdot B = 0}$).  

\begin{figure}[htbp]
\centering
\includegraphics[width=\columnwidth]{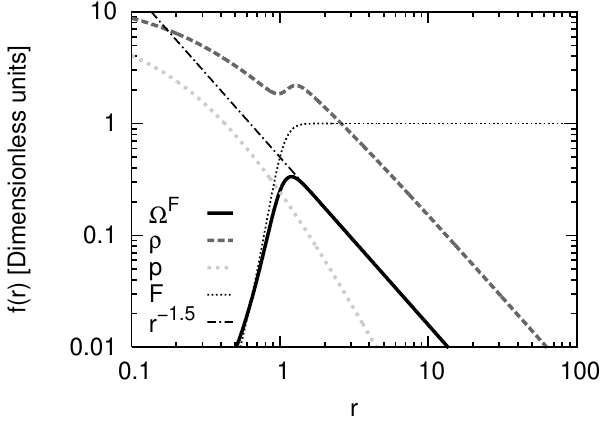}
\caption{
Profiles of the fixed in time variables for the inlet in hydrodynamic equilibrium.  Here we give constraints on $\Omega^F,\rho,p,E_{\phi}$ ($E_{\phi}=0$ not shown).  
Parameters are $v_{\rm K}=0.5,\epsilon=2/3$. 
The thin dotted line is the Fermi step function used to smooth those variables experiencing a sharp transition at the inner disk radius $r=1$.  
\label{fig:disk}}
\end{figure}

By letting the jet solution alone determine Poynting- and mass- flux, 
we loose control over the energy flux parameter $\mu$ and the limiting asymptotic 
Lorentz factor $\Gamma^*$.  
It will rather be a consequence of the MHD under the constraints we have given, while we have used our freedom to provide a boundary most closely resembling a realistic hot disk corona.
A graphical summary of the disk wind boundary conditions is shown in Fig.\,\ref{fig:disk}. 

In order to extend the parameter space towards higher $\mu$, we also investigate
cases where we have over-determined the boundary conditions by specifying the mass-flux 
through
\eqon
v_{z}(r,0) = v_{\rm inj} v_{\phi}(r,0)
\eqoff
similar to e.g. \citet{1997ApJ...482..712O, 2002A&A...395.1045F}.
Another parameter run adopts a $1/r$ profile for the toroidal 
magnetic field $B_{\phi}(r) = - \eta F(r) / r $ in order 
to specify the Poynting flux with an additional parameter $\eta$.  
These runs 
\red
and the influence of the over-determination 
\black
is discussed separately in section \ref{sec:extended}.

\subsubsection{Outflow boundaries ($R_{end},Z_{end}$)}
The standard outflow boundary conditions for many numerical codes are zero-gradient 
conditions, which are usually sufficient as the plasma velocities are constrained to be outward-pointing.  

\red
In the case of sub fast-magnetosonic outflows, this strategy is unfortunately insufficient as the flow inside of the domain will depend on the flow beyond the boundary via the incoming characteristics.  
Just as for the inlet boundary, the now missing information has to be supplied by constraints that describe best the physical conditions downstream of the boundary.  
In the case of an outflow, the conditions leading to an untampered flow are however impossible to know a priori.  
A way to circumvent this unphysical feedback is to avoid any causal contact by moving the boundary far away such that the characteristics will not enter the domain of interest within the simulated time.  
\black

\red
When considering a boundary outside of causal contact ``very far away'', we estimate for Alfv\'en waves to travel over $10^3$ scale radii within the anticipated simulation time.  The computational effort of such huge grids does not allow a large parameter study at the current time and we must leave this option for future endeavours.  
\black

In the absence of a substantially better solution, zero-gradients are  used for the 
primitive variables except for magnetic fields for which this simple approach leads 
to artificial electric currents implying an inward-pointed Lorentz-force.  
Especially for low plasma-$\beta$ this may result in a devastating artificial
collimation - preventing any steady state to establish and artificially collimating 
the outflow increasingly thin with time. 

\cite{1999ApJ...516..221U} have performed a systematic study comparing different 
approaches for outflow conditions including a (toroidal) force-free condition 
$\mathbf{j_{p}||B_{p} = 0}$ and a more sophisticated version including an additional numerical factor that needs to be determined a posteriori.  
For the outflow conditions in our simulations we instead recover the magnetic field 
components by imposing constraints on the poloidal 
($j_{r}=-\partial_{z} B_{\phi}$, $j_{z}=r^{-1}\partial_{r}rB_{\phi}$) and
toroidal $(j_{\phi}=\partial_{z}B_{r}-\partial_{r}B_{z})$ electric  currents.  
For the toroidal magnetic field (poloidal electric current) we radially extrapolate 
the expected $1/r$ law of a marginal $j_{z}$ at the radial end ($\rm R_{end})$.
Using $\partial_{z}B_{\phi}=0$ allows to specify $j_{r}$ at the upper end of the domain ($\rm Z_{end})$.  
Concerning the poloidal magnetic field components we implement a current-free 
boundary condition by enforcing $j_{\phi} = 0$.
This is a novel approach designed to minimize spurious effects of collimation.
\red
We convinced oursevels that boundary effects have only a marginal effect on the solution by varying the grid-size and geometry.  
For a detailed discussion and comparison of various outflow conditions we refer to appendix \ref{sec:zerocurrent}.  
\black  
 
We note that, as a further complication, a fully relativistic version for a force-free
or force-balance boundary conditions would also need to take into account electric 
forces. 
We have estimated the impact of such an upgrade and found that due to the geometry
of our outflow (in particular the location of the light surface)
it would play a minor role and is thus not worth the effort to implement.

\subsection{Initial conditions}
As initial state we prescribe a force-free coronal magnetic field,
$F^{\alpha \beta}j_{\beta} = 0$, together with a gas distribution in hydrostatic
equilibrium.
Both is essential in order to avoid artificial relaxation processes
caused by a non-equilibrium initial condition.
We apply a polytropic equation of state $p=K\rho^\gamma$ with a ``classical''
polytropic index of $\gamma=5/3$ since our flows are always cold when compared
to the rest-mass.  
To further strengthen this choice, we performed a comparison simulation with the 
\cite{1948PhRv...74..328T} equation of state as described by \cite{Mignone2005} 
which produced an identical jet once the hot shock has passed through.  
The constant $K$ is determined by the radial force-balance of the inlet.  

For the initial magnetic field configuration we apply two different geometries.
\red
	Field configuration A
\black
is a potential field of hourglass shape as applied by 
\citet{1997ApJ...482..712O} and \citet{2002A&A...395.1045F} with the magnetic field components
\begin{eqnarray}
B_{r} &=& \frac{1}{r} 
	\left[1-\frac{(z+z_{d})}{\left(r^2+(z+z_{\rm d})^2\right)^{1/2}}\right]\\
B_{z} &=& \frac{1}{\left(r^2+(z+z_{\rm d})^2\right)^{1/2}},
\end{eqnarray}
corresponding to a vector potential 
\begin{equation}
A_{\phi} =\frac{1}{r} \left[\sqrt{r^2+(z+z_{d})^2}-(z+z_{\rm d})\right]\label{eq:aphi-op}
\end{equation}
in cylindrical coordinates with $B_{r} = -\partial_{z} A_{\phi}$ and
$B_{z} = r^{-1}\partial_{r} \ r A_{\phi}$.
The dimensionless disk thickness $z_{\rm d}$ with $(z_{\rm d}+z)>0$ is introduced to
avoid kinks in the field distribution for $z<0$ (the ghost zones) and we choose $z_{d}=1$ for convenience.  

Our other option for the initial magnetic field 
\red
(configuration B)
\black
is the ``split monopole'' 
\citep{1987PASJ...39..821S} with the magnetic field components  
\begin{eqnarray}
B_{r} &=& \frac{r}{\left(r^2+(z+z_{d})^2\right)^{3/2}}\\
B_{z} &=& \frac{z+z_{d}}{\left(r^2+(z+z_{d})^2\right)^{3/2}}.
\end{eqnarray}
In the split-monopole setup 
the parameter $z_{\rm d}$ defines the offset of the 
fiducial center of the monopole from the grid origin, and adjusts 
the initial angle of the field lines with respect to the disk-surface.
We either adopt an angle of $\theta=85^\circ$ or $\theta=77^\circ$ for the field line passing through $r=1$.
Similar to Eq. \ref{eq:aphi-op}, the split monopole field can be described by a vector potential
\begin{equation}
A_{\phi} = \frac{1}{r}
\left[1 - \frac{z+z_{\rm d}} {\left(r^2+(z+z_{\rm d})^2\right)^{1/2}}\right].
\end{equation}

The fields are scaled to satisfy the the choice of the plasma-$\beta$
\eqon
\beta \equiv \left.\frac{B_{p}^2}{8\pi p}\right|_{r=1,z=0}
\eqoff 
at the inner disk radius.  

\red
	It should be kept in mind that plasma-$\beta$ largely varies along the disk boundary.  In configuration A, the profile $\beta(r)$ monotonically decreases until for large radii it is $\beta(r)\propto r^{-0.5}$ leading to a magnetically dominated outer corona.  
	In the split-monopole, $\beta(r)$ decreases first to a minimum value (at $r^*(\theta=77^{\circ})\approx5$ and $r^*(\theta=85^{\circ})\approx15$) and increases for large radii according to $\beta(r)\propto r^{1.5}$ leading to thermal dominance.  
\black

In summary, for our injection boundary condition we are left with the following five dynamical parameters,
\begin{equation}
(v_{\rm K},\beta,\epsilon, v_{\rm inj},\eta),
\end{equation}
where strictly speaking we are only allowed to choose the first three when launching 
sub-slow. 
An overview of the simulations performed in this parameterization is shown in 
Tab.\,\ref{tab_all}.  

\subsection{Numerical grid and physical scaling}

We use a numerical grid of $512\times1024$ cells applying cylindrical coordinates.
Onward from the inner region $(r<1,z<1)$, which is resolved with $20\times20$ 
equidistant cells, we apply a stretched grid with the element size increasing 
by a factor of $\lesssim 1.005$. 
This leads to a domain size of $(r\times z)=(100\times200)r_{\rm i}$ 
corresponding to $(300\times600)\ r_{\rm s}$ if $r_{\rm i}=3r_{\rm s}$ (see sketch in figure \ref{fig:setup}).  
Staggered magnetic fields treated via constrained transport \citep{1999JCoPh.149..270B} are used to ensure $\mathbf{\nabla\cdot B = 0}$.

Because of the constraints imposed on the cell aspect ratio by the zero-current 
boundary (appendix \ref{sec:zerocurrent}), 
we set the last five grid cells to be equally spaced with maximal aspect 
ratios $<3/1$.  

\red
The dimensionless nature of our simulations allows for various astrophysical interpretations.  We provide a physical scaling of simulation variables (marked with a prime) in the following paragraph.  

Since velocities are given in terms of the speed of light ($c'=1$), relativistic simulations are in need of only two additional scales.    
The simulation variables are connected to their physical counterparts via
\begin{eqnarray}
v&=&v'c;\ \ l=l'l_{0}; \ \ t=t't_{0}=t'l_{0}/v_{0}; \ \ \rho = \rho ' \rho_{0} \\
p &=& p'p_{0} = p' \rho_{0}c^2; \ \ B = B'B_{0} = B'\sqrt{4\pi\rho_{0}c^2}.
\end{eqnarray}
If we assume a Schwarzschild black hole as central body, we may set the spatial scale $l_{0}=6r_{g}$, equating the inner disk radius with the ISCO.  Then it becomes
\begin{eqnarray}
v_{0} &=& 3\times 10^{10} {\rm cm}\ {\rm s}^{-1}\\
l_{0} &=& 9\times 10^{5}{\rm cm} 
                       \left(\frac{\rm M_{\bullet}}{\rm M_{\odot}}\right)\\
t_{0} &=& 3\times 10^{-5} {\rm s} 
                       \left(\frac{\rm M_{\bullet}}{\rm M_{\odot}}\right).
\end{eqnarray}
Assuming a physical outflow mass-loss rate in terms of the Eddington limited accretion rate
$\dot{M}=0.01\dot{M}_{\rm edd}$ 
we can provide a scale for the density by comparison to the mass loss rate of the simulation $\dot{M}'$
\eqon
\rho_{0} = 6\times 10^{-7}\frac{1}{\dot{M}'}\left(\frac{M_{\bullet}}{M_{\odot}}\right)^{-1}\rm g\ cm^{-3}
\eqoff
where we applied a radiative efficiency of $\eta^*=0.1$.  
The scaling of pressure and magnetic fields then follows as 

\begin{eqnarray}
p_{0} &=& 5\times 10^{14}\frac{1}{\dot{M}'}\left(\frac{M_{\bullet}}{M_{\odot}}\right)^{-1}\rm g\ cm^{-1}\ s^{-2} \\
B_{0} &=& 8\times 10^7\left.\dot{M}'\right.^{-0.5}\left(\frac{M_{\bullet}}{M_{\odot}}\right)^{-0.5} \rm Gauss . 
\end{eqnarray}

Under these considerations, the only remaining scaling parameter is the mass of the compact object $M_{\bullet}$.  
Neglecting additional physical processes as radiation pressure or radiative 
cooling leaves us with a scale-free model that can be applied to any disk-wind
launched jet around compact objects.  
Table \ref{tab:scaling} provides a fiducial scaling for a microquasar with $M_{\bullet}=10 M_{\odot}$ and for an AGN with $M_{\bullet}=10^8 M_{\odot}$.  
\black
The scale-free nature becomes obvious if we recall the rest-frame temperatures
for an ideal gas,
$
        T = \frac{p'\ c^2}{\rho' k_{\rm B}}\langle \mu \rangle \rm m_{\rm p}
$
\citep{1989cup..book.....A},
with the mean molecular weight $\langle \mu\rangle$,
    the proton mass ${\rm m_{\rm p}}$,
and the Boltzmann constant $k_{\rm B}$.
For ionized hydrogen $\langle\mu\rangle=0.5$ one would find temperatures of
$T = (p'/\rho')\ 5.45\times10^{12} \rm K$,
while for an electron-positron plasma ($\mu\simeq1/2000$) the temperatures are 
lower by three orders of magnitude.
These ultra-high scaling temperatures are only reached in the very inner corona between
the central object and the inner disk radius. 
As we do not intend to follow the dynamics here, this does not really pose a problem. 
In principle, once $p'/\rho'\gtrsim 1$, the
equation of state transcends towards $\gamma = 4/3$ according to a Synge-gas.
In the jet, thermal pressure quickly looses importance and the temperatures 
are significantly lower.

\begin{deluxetable}{lllllll}
\tablewidth{\columnwidth}
\tabletypesize{\scriptsize}
\tablecaption{Fiducial scaling\label{tab:scaling}
}
\tablehead{
\multirow{2}*{ $\frac{M_{\bullet}}{M_{\odot}}$  }&
$l/l'$ & 
$t/t'$ &
$\rho/\rho'$& 
$p/p'$ &
$B/B'$ \\
& 
$\rm [cm]$ &
$\rm [s]$ &
$\rm [g\, cm^{-3}]$ &
$\rm [g\, s^{-2}\, cm^{-1}]$ &
[Gauss]
}
\startdata
$10^8$ & 
$9\times10^{13}$&
$3\times 10^{3}$ &
$1.8\times10^{-16}$  &
$1.5\times10^{5}$  &
$1.4\times 10^3$\\
$10$ &
$9\times10^{6}$ &
$3\times10^{-4}$&
$1.8\times10^{-9}$ &
$1.5\times10^{12} $&
$4.4 \times 10^{6}$
\enddata
\tablecomments{Scaling for simulation WA05 ($\dot{M}'=32.67$) assuming a physical mass loss rate of $\dot{M}=1\%\dot{M}_{edd}$ with an efficiency of $\eta*=0.1$.  }
\end{deluxetable}
%

%
\section{Results and discussion}\label{sec:results}

We now present the results of our numerical simulations considering the formation
of relativistic MHD jets from accretion disks.
Each simulation consumed approximately 48 hours on 16 processors.  
The overall goal is to test whether the paradigm of MHD self-collimation
of non-relativistic jets established from numerical simulations
\cite{1995ApJ...439L..39U, 1997ApJ...482..712O, 1999ApJ...526..631K, 2002A&A...395.1045F}
also holds in the relativistic case.

%

\begin{deluxetable*}{ccccccl||crrrrr}
\tablecaption{Parameter summary of our disk-wind simulations. \label{tab_all}
}
\tabletypesize{\scriptsize}
\tablewidth{\textwidth}
\tablehead{
 \colhead{ID}                      & 
 \colhead{Top}               &
 \colhead{$\beta$}                     & 
 \colhead{$\epsilon$}       & 
 \colhead{$v_{\rm inj} $}        & 
 \colhead{$\eta$}         &
 \colhead{Remarks}      &
 \colhead{$\Gamma_{\rm max}$} &
 \colhead{$\mu_{\rm max}$} &
  \colhead{$\xi$} &
 \colhead{$v_{p,\rm max}$} &
 \colhead{$r_{\rm jet}$} &
 \colhead{$\dot{M}$}
 }
\startdata
%







WA01 & A & 0.2 & 2/3 & var. & var. & - & 
              1.23&         1.33 &        18.84 &       0.54 &        20.26 &        23.77
\\

WA02 & A & 1 & 2/3 & var. & var. & - & 
       1.27 &         1.37 &        11.47 &       0.58 &        21.86 &        26.62
\\
WA03 & A & 2 & 2/3 & var. & var. & - & 

1.26&         1.34 &        10.69 &       0.57 & 
       22.21 &        24.15
\\
WB01 & B & 1 & 2/3 & var. & var. & $\theta=77^\circ$ & 
       1.33&         1.41 &        8.27 &       0.64 &        24.19 &        16.48
\\
WB02 & B & 1 & 2/3 & var. & var. & $\theta=85^\circ$ & 
1.29&         1.33 &        8.19 &       0.60 & 
       21.27 &        15.66
\\
WA04 & A & 0.2 & 1/6 & var. & var. & - & 
       1.27&         1.38 &        13.40 &       0.58 &        21.14 &        36.08
\\
WA05 & A & 1 & 1/6 & var. & var. & - &
1.25&         1.33 &        9.84 &       0.57 &         22.77 &        32.67
\\
WA06 & A & 2 & 1/6 & var. & var. & - &
1.25&         1.32 &        9.34 &       0.57 & 
       22.92 &        29.27

%





\enddata
\tablecomments{Columns are from left to right: simulation ID; 
 initial magnetic field distribution (Top): potential field (A) or 
     split monopole (B);
 plasma-$\beta$; accretion disk temperature parameter $\epsilon$; 
     injection speed $v_{\rm inj}$ as a fraction of $v_{\phi}(r)$;
     toroidal disk-field scaling $\eta$;  specific remarks; to the right we show values of the steady state, the maximum Lorentz factor $\Gamma_{\rm max}$ collimation degree $\xi$, maximal poloidal velocity $v_{p,\rm max}$, jet radius $r_{\rm jet}$ and the total mass flux out of the domain $\dot{M}$. 
}
\end{deluxetable*}


\subsection{Overall evolution of the outflow}

The initial evolution of the disk corona is governed by the propagation of 
toroidal Alfv\'en waves launched due to the rotation of the field line 
foot-points.  
The initial force-free magnetic field structure is adapted to a new dynamic
equilibrium according to a rotating wind magnetosphere.

\begin{figure*}[htbp]
\centering
\includegraphics[width=0.7\textwidth]{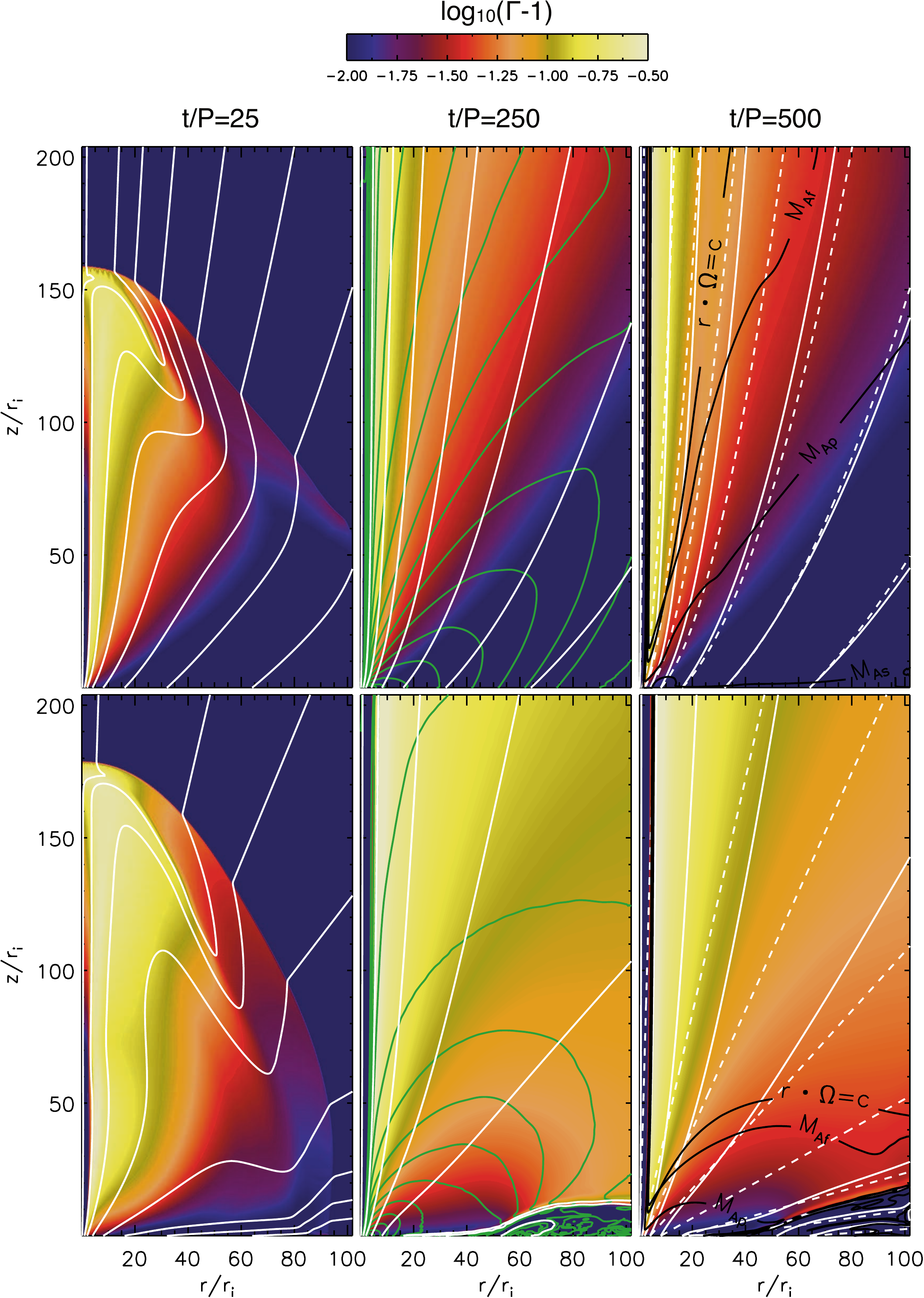}


%


\caption{
Formation of a relativistic MHD jet.
Shown is the Lorentz factor (color gradient) in terms of $log_{10}(\Gamma-1)$ 
at the time of 25, 250 and 500 (left to right) inner disk rotations. 
{\em Top:} Field distribution case A, hourglass-shape potential field 
(run WA02).
{\em Bottom:} Split monopole initial field distribution case B (run WB01). 
Shown are poloidal magnetic field lines (solid white lines);
the critical MHD surfaces (solid black lines);
the light surface $r\cdot \Omega = c$ (solid black line). 
For time $T/P=250$ electric current flow lines are added (solid green line).
Time step $T/P=500$ also show the initial field lines for comparison 
(dashed white lines).
\label{fig:simul}}
\end{figure*}

A wind is launched from the disk boundary and is continuously accelerated 
driving a shock front through the initial hydrostatic corona and sweeping
this material out of the computational domain (Fig.\,\ref{fig:simul}).
The disk wind evolves into a collimated outflow of super-magnetosonic speed.
Along the symmetry axis the hydrostatic initial condition is very well 
preserved.
Once the bow shock has passed through the domain, the jet mass flux declines
to a value which is solely governed by the internal outflow dynamics
and the injection boundary conditions. 
Similarly, the post-shock magnetic field distribution follows as well
from the internal outflow dynamics and has in principle little in common with 
the initial setup.  
%
Certain combinations of boundary conditions for mass flux and magnetic field
will result in a quasi-stationary\footnote{We denote the dynamical state as
{\em quasi} stationary as due to Keplerian disk rotation the disk outflow
a large radii has evolved for a considerably lower number of disk rotations.
Therefore, a slight change in the dynamical state of the outer outflow
can be expected after another few outer, thus 1000s of inner rotations.}
state of the outflow evolution 
(see next section).
From this point onwards we can start our investigations of collimation and
acceleration.
In this paper we concentrate on analysis when the flow has reached a
quasi-steady state.
We usually terminate our simulations after 500 inner disk-rotations $P$, 
while a quasi-steady state is established over most of the domain after 
about 200 rotations.

Figure \ref{fig:simul} shows the time evolution for two exemplary 
simulations with an initial hourglass-shaped potential field distribution
(case A) and a split-monopole field distribution (case B),
each for the parameter choice $(\beta,v_{K},\epsilon) = (1,0.5,2/3)$.
%

The figure shows the 
Lorentz factor, the poloidal magnetic field lines, poloidal electric current flow lines and the critical MHD surfaces.
In addition the light surface
is drawn.

Phenomenologically, the solutions form a magnetic nozzle with, depending on the disk flux distribution, considerable difference in the width, but comparable final opening angles of the fast component.  A broader initial field distribution (case B) also results in broader and faster winds where the material originating from the inner disk is more effectively thinned out.  
In analogy to hydrodynamic nozzles, the flow reaches the slow-magnetosonic speed directly above the throat.  
Collimation happens mainly before the fast-magnetosonic surface is reached. Afterwards, the opening angle of a given field line is approximately conserved.  

Of particular interest is the electric current distribution
(shown for the time step $T/P = 250$).
The electric current distribution 
is a consequence of the dynamical
evolution of the outflow and therefore a direct outcome of the
disk boundary magnetic flux profile and the Keplerian field line rotation.

In general, the electric current leaves the outer disk to return within the fast component of the outflow.  
It is expected to enter the inner disk and then flow radially outwards 
closing with the outgoing current.
%
Such butterfly-shaped circuits are expected in Keplerian disks while
the $j_{r}$ plays a leading role 
in the disk-jet feedback \citep{1997A&A...319..340F}.
A positive radial electric current in the disk-corona supports accretion 
by braking the disk material due to its magnetic torque $j_{r}\times B_{z}$ 
similar to a Barlow-wheel\footnote{
However, this region is not resolved 
within our numerical domain, as it is located below our injection boundary
as part of the underlying non-ideal MHD accretion disk. 
See \cite{2002ApJ...581..988C, 2007A&A...469..811Z} for non-relativistic
simulations of the disk-jet interaction}.

The inclination between the poloidal current vector and the magnetic field line
indicates the direction of (de-) collimating magnetic forces acting on the flow.
When the inclination becomes less than $90^{\circ}$, the Lorentz force
$j_{p}\times B_{\phi}$ changes from collimation to de-collimation.  
This can be clearly seen in the snapshots at $T/P=250$ of the case A simulation where actual
field lines are indicated in white and initial field lines in red.
In the actual field distribution, the field lines are somewhat pushed away from
the surface $\mathbf{j_{p} \perp B_{p}}$ (this is also where magnetic acceleration is most effective).
%
For case B this happens  
beyond the light surface.
As a result, both electric and magnetic forces deflect the flow towards 
the disk boundary which leads to a highly unstable layer just above 
the outer disk.
We will provide an in-depth analysis including all forces acting on the flow
in section \ref{sec:trans-field}.

The locations of the characteristic surfaces are signatures for the MHD flow.
Depending on the initial magnetic flux distribution (case A,B) and the 
mass flux profile (see also \citet{2006ApJ...651..272F}), 
this location may vary a great deal. 
In our case B simulations, we generally observe surfaces which leave the domain
in radial direction (parallel to the disk surface).
For the case A simulations these surfaces tend to "collimate" leaving the domain
in vertical direction. 
The latter implies a two-layered structure of the jet - a central super-fast magnetosonic jet surrounded by a 
sub-Alfv\'enic outflow.
This is an interesting aspect for observational modeling 
and for stability analysis of sheath-spine jets 
\citep{2005MNRAS.356..859P, 2007ApJ...662..835M, 2007ApJ...668L..27K, 
       2007ApJ...664...26H, 2009MNRAS.397.1486B}.
The broad wind launched from the outer regions of the disk has much lower velocities, decreasing continously with increasing launching-radius. For example, the terminal velocity of the flow originating from $r_{\rm fp}>32$ of the case A simulations drops below $0.2c$, consistent with the X-ray absorption features observed in a mounting number of AGN
\citep{2006AN....327.1012C, 2009A&ARv..17...47T}.  
In principle, our dynamical models can provide basic ingredients (e.g. flow geometries and velocity gradients) for the modeling of spectral line profiles of disk winds \citep{1995MNRAS.273..225K, 2008MNRAS.388..611S}.

\begin{figure}[htbp]
\centering
\includegraphics[width=\columnwidth]{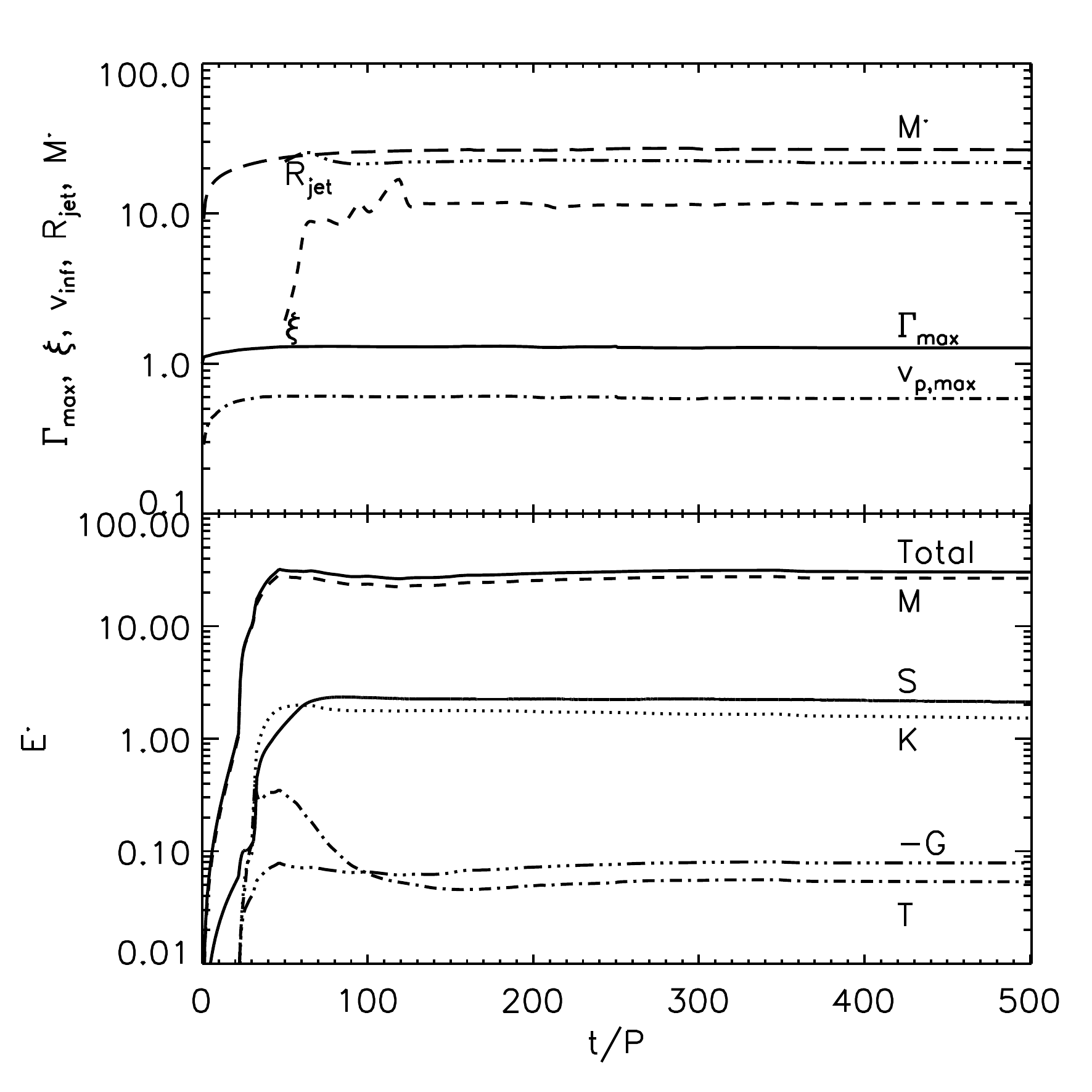}
\caption{
\red
Time evolution of characteristic quantities.  \textit{Top panel:} After an initial adjustment till $t/P\simeq200$, mass flux $\dot{M}$, jet radius $r_{jet}$, collimation degree $\xi$, maximal Lorentz-factor $\Gamma_{\rm max}$  and poloidal velocity  $v_{p, \rm max}$ cease to evolve. 
\textit{Lower panel}: Power in the individual energy chanels out of  the domain. Thermal power ($T$)
peaks when the shock reaches the upper boundary and is negligible otherwise.
The total outgoing power (labeled accordingly) is dominated by rest-mass ($M$) and Poynting flux ($S$). Also shown are the gravitational ($G$) and (purely) kinetic ($K$) contributions (simulation run WA02).  
\black
\label{fig:diag}}
\end{figure}

Following  \cite{2006ApJ...651..272F}, we may define an average collimation 
degree $\xi$ of the outflow measured as the fraction of vertical and radial 
mass flux through equal-area surfaces at a certain height (here at $z=z_{m}$),
\begin{equation}	
	\xi = \frac{
	\int_{0}^{r_{m}}r \Gamma v_{z} \rho |_{z_{m}}\ dr
	}{
	\int_{z_{m}-r_{m}/2}^{z_{m}}r_{m}\Gamma v_{r}\rho |_{r_{m}}\ dz
	}.
\end{equation}
Similarly, we define a mass flux weighted jet radius,
\begin{equation}
	r_{jet} = \frac{
		\int_{0}^{r_{m}}r \Gamma v_{z} \rho |_{z_{m}} \ dr
	}{
	\int_{0}^{r_{m}} \Gamma v_{z} \rho |_{z_{m}}\ dr
	}.
\end{equation}
The corresponding values for $\xi$ and $r_{\rm jet}$ 
derived for $z_{m}=200$ at the upper end of 
the domain and for time $t/P=500$ are given in Tab.\,\ref{tab_all}
along with the maximum Lorentz factor $\Gamma_{\rm max}$, the maximum
poloidal velocity $v_{p,max}$, and the total mass flux $\dot{M}$.  
Figure \ref{fig:diag} shows the time evolution of these quantities in the top panel.  In general, we observe that the collimation degree $\xi$ is the most sensitive tracer for secular trends among the observables mentioned.  
In the lower panel, we show the evolution of jet-power in the individual energy channels leaving the computational domain (radial and vertical).  
After the re-configuration of the initial stationary-state to the dynamical solution, the partitioning of energies is completed at around $t/P=100$.  Thermal energy-flux peaks when the hot bow-shock passes through the upper boundary.  Far away from the central object, gravitational and thermal energy-flux are negligible 
The integrated energy-flux is dominated by rest-mass, reflecting the fact that only the inner component reaches significant Lorentz-factors.  
The balance between Poynting and kinetic flux is of particular interest. Figure \ref{fig:diag} shows merely the end result of the spatial conversion history with the remaining electromagnetic energy $\mathcal{S}$ above the purely kinetic part $\mathcal{K}$.  
More detailed insight into how this is established is provided in the following section using an individual field line. 
\subsection{Stationary state analysis}
Simulations starting from an initial field distribution A evolve into
a quasi-stationary flow solution after about 200 inner disk rotations.
Figure \ref{fig:stationary_flow} shows our reference simulation WA04
at time $t/P=250$, including an enlarged subgrid of the innermost area
of the domain.

Steady state solutions are helpful to understand the flow structure
for a number of reasons. 
Firstly, by using MHD conservation laws, the conserved quantities 
(see Sect.\,\ref{sec:flconst})
allow to identify the momentum and {\em energy channels} of the flow during 
acceleration and collimation.
Secondly, by using the force-balance equations 
\ref{eq:trans-field},\ref{eq:parallel-field} we may identify the
leading forces on the material along the outflow.
Thirdly, the cross-check for conserved quantities provides another test 
for the quality of our setup and the numerical approach.
A secondary indicator of stationarity is the alignment of poloidal velocities
with the poloidal magnetic field lines, $E_{\phi} = 0$.
Figure \ref{fig:stationary_flow} shows corresponding velocity vectors
confirming this picture.

\begin{figure}[htbp]
\centering
\includegraphics[width=\columnwidth]{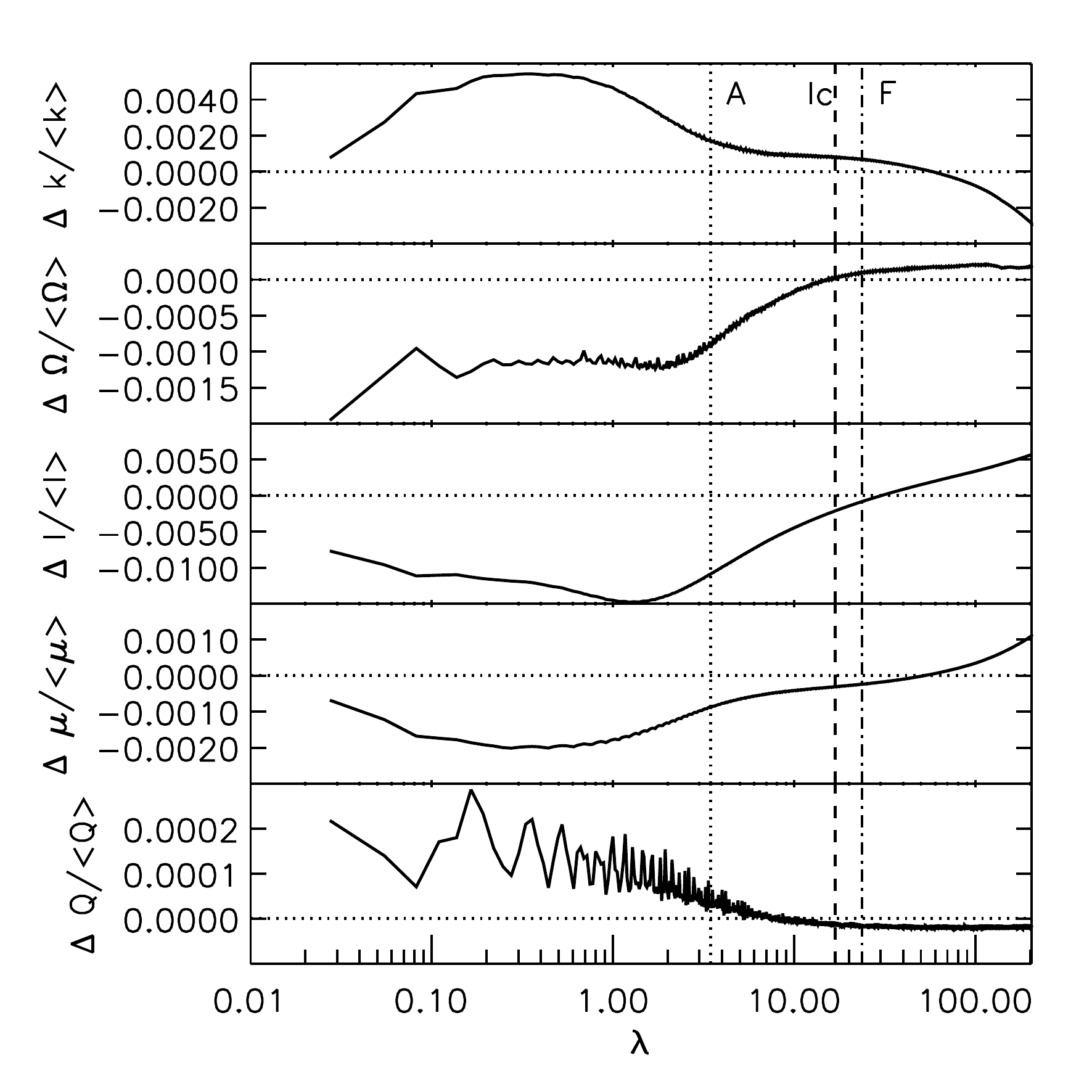}
\caption{Field line constants (conserved quantities).
Shown is the deviation from the average value along the field line
rooted at $r_{\rm fp}=2$ (simulation run WA02).
\label{fig:const}}
\end{figure}

\begin{figure}[htbp]
\centering
\includegraphics[width=8cm]{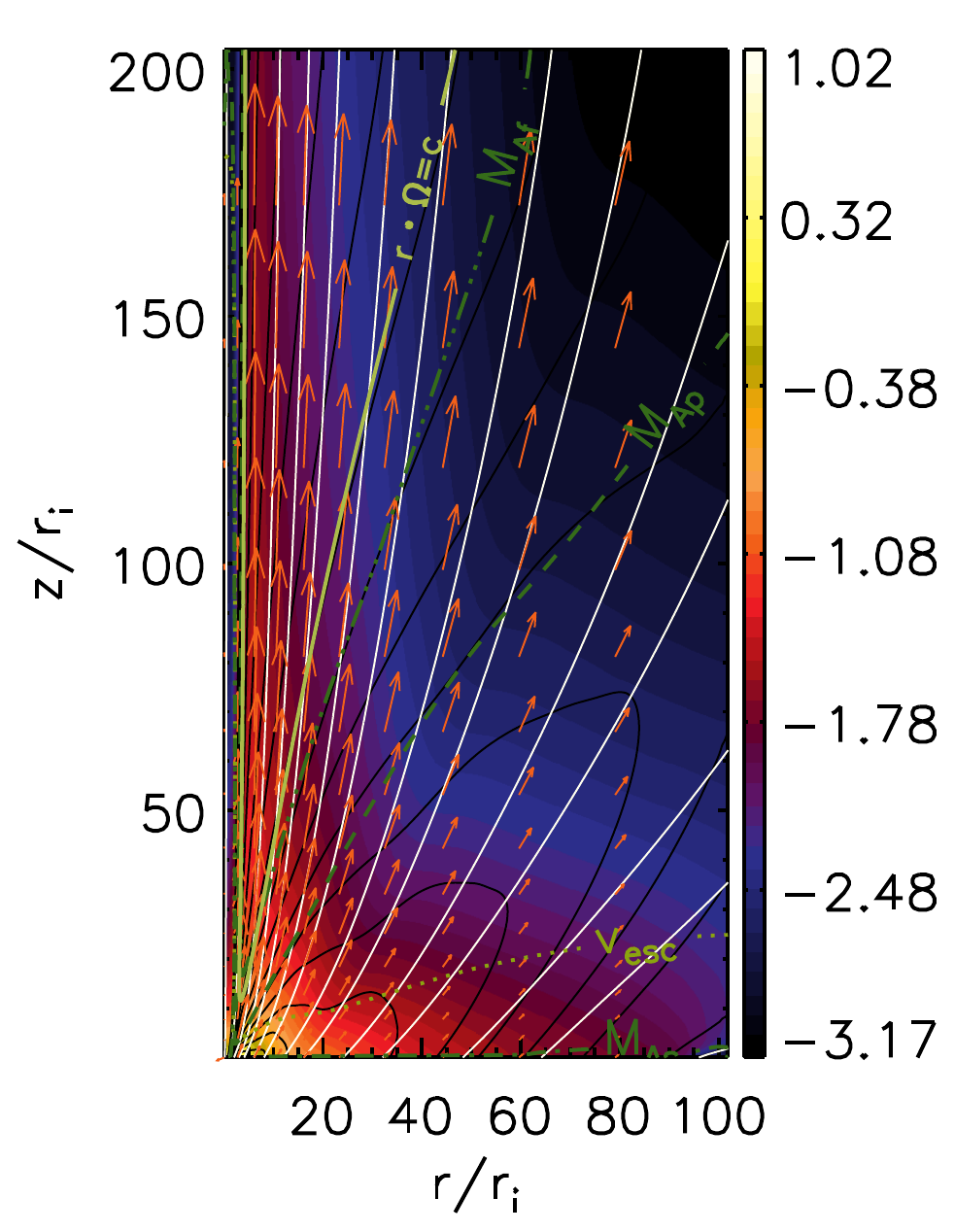}
\includegraphics[width=8cm]{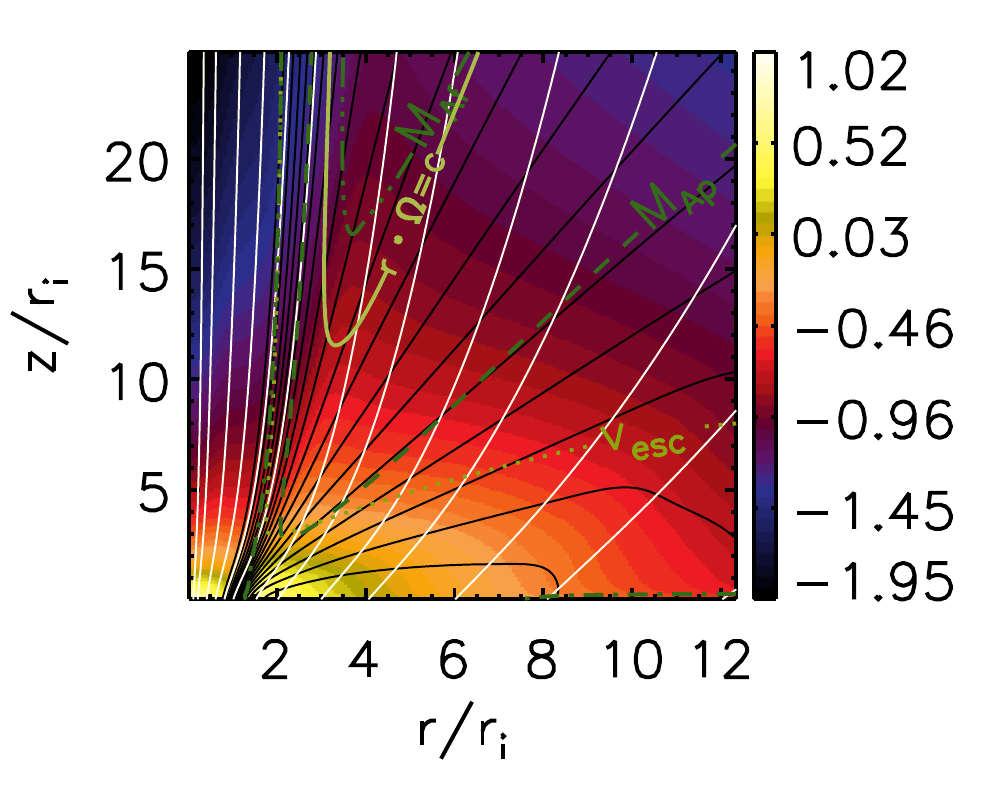}
\caption{Logarithmic (rest-frame) density of the stationary 
flow (simulation run WA04).
Shown are poloidal magnetic field lines (solid white), 
electric current flow lines (solid black), 
characteristic MHD surfaces (various dot-dashed green), 
surface of escape velocity (dotted green), 
light surface (solid green).  
Arrows in the top plot indicate the velocity field.
The bottom figure is an enlarged picture of the central region 
indicating the three regimes defined by the light surface.  
\label{fig:stationary_flow}}
\end{figure}

This is confirmed by checking in detail the complete set of integrals of motion
of the MHD-flow $k,\Omega^F,Q,l,\mu$ as defined in equations $\ref{eq:kf}-\ref{eq:mu}$. 
Figure \ref{fig:const} shows the relative deviation of these quantities
from their average value along a given field line after $t/P=500$.  
The integrals are conserved within $1\%$-accuracy already right above the 
injection boundary - clearly demonstrating the quality the choice 
of our numerical setup, in particular the injection boundary conditions
carefully constructed from an equilibrium of Keplerian rotation and 
gas pressure.

Due to the differential rotation law, the number of Keplerian rotations 
$t/P(r)$ scales with radius as $t/P(r) = (t/P) r^{-3/2}$, implying
that at the end of our simulations ($t/P=500$), 
we have performed roughly one rotation at $r=64$ and half a rotation 
at $r=100$. 
Nonetheless the integrals of motion for the field line $r_{\rm fp}=64$ are 
conserved within $0.1\%$.

\subsubsection{Collimating and accelerating forces}\label{sec:trans-field}
In this section we identify the forces responsible for jet acceleration
and collimation applying the steady-state parallel and transversal
force-equilibrium Eqs.\,\ref{eq:trans-field},\ref{eq:parallel-field}.

\begin{figure*}[htbp]
\centering
\includegraphics[width=0.49\textwidth]{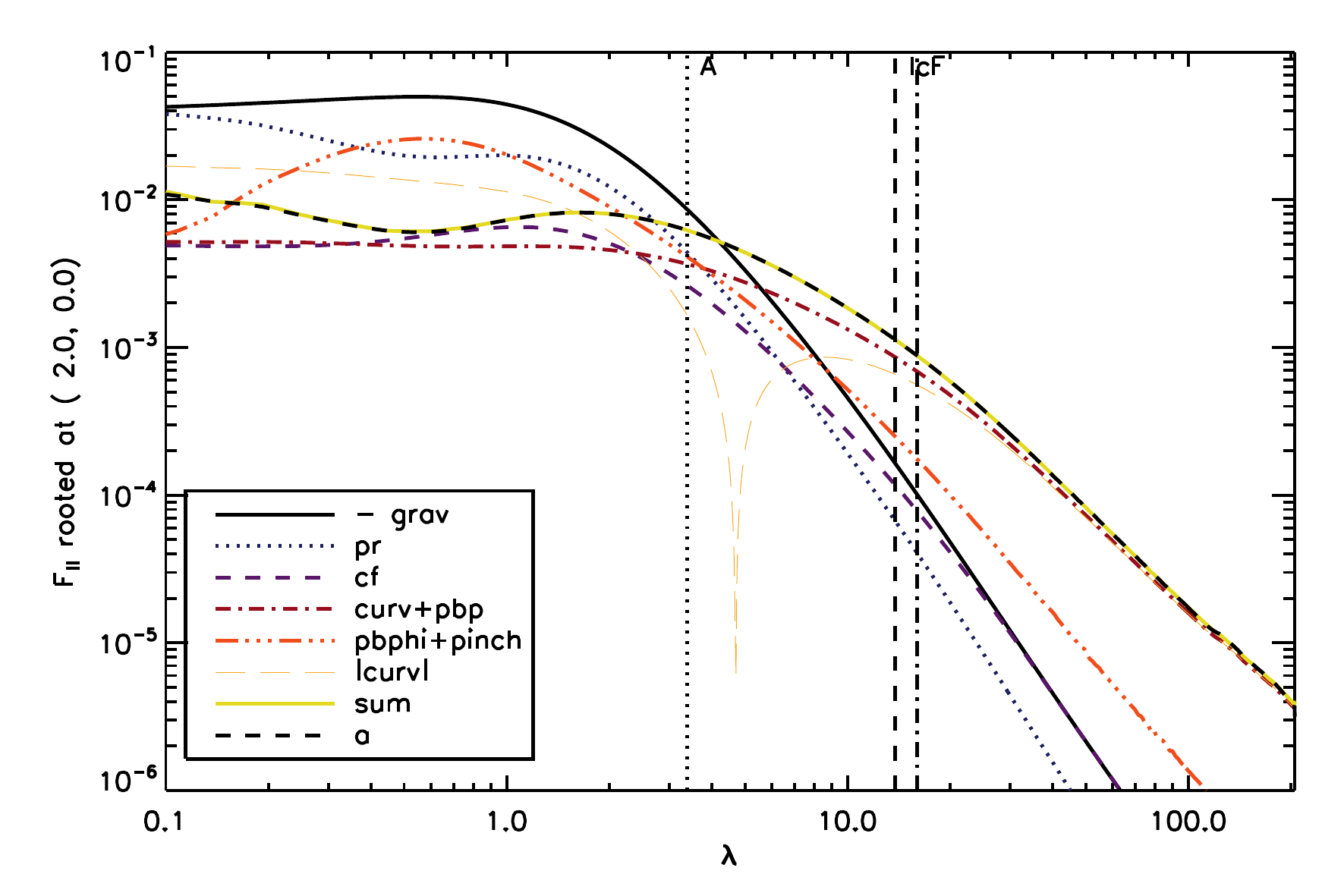}
\includegraphics[width=0.49\textwidth]{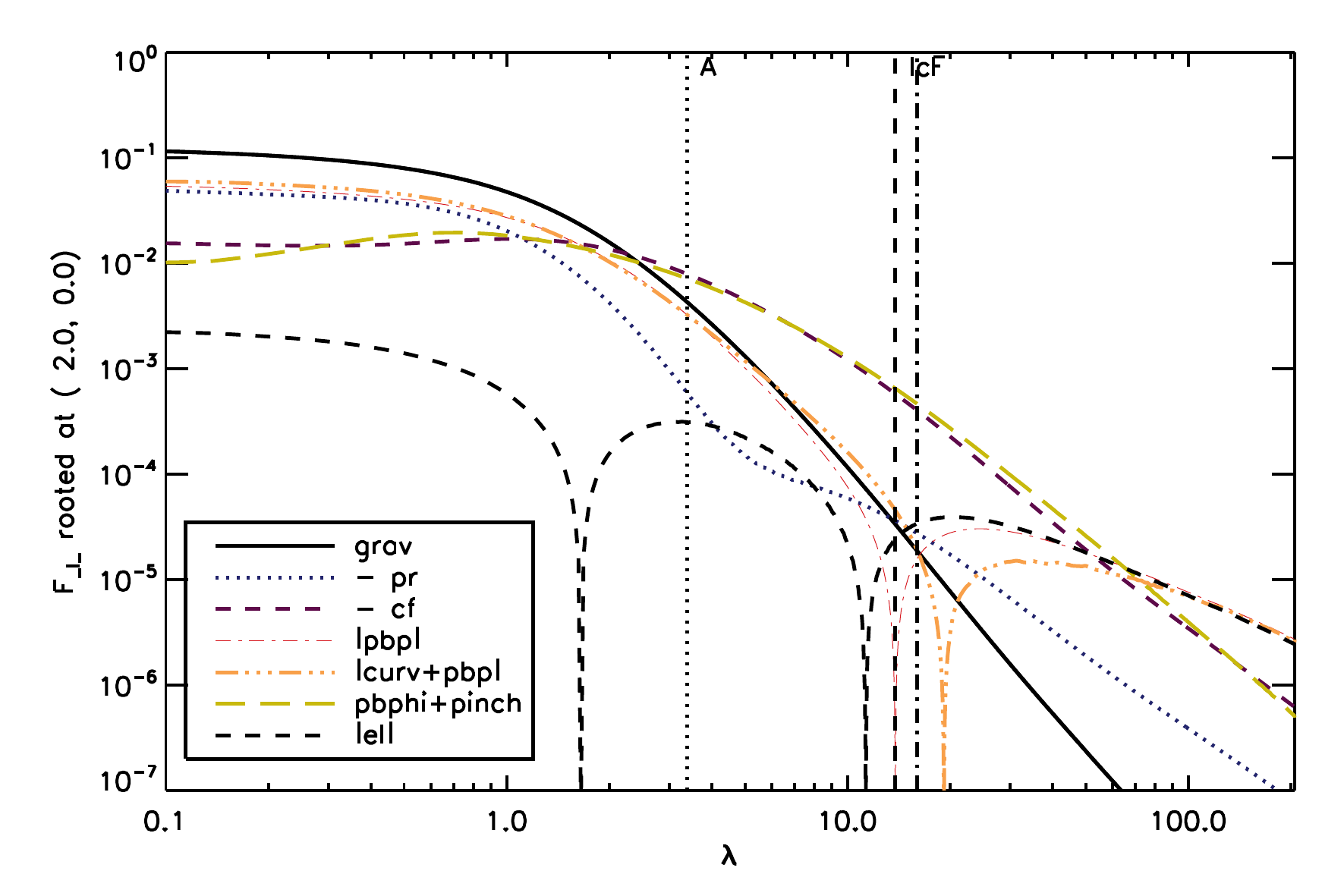}
\includegraphics[width=0.49\textwidth]{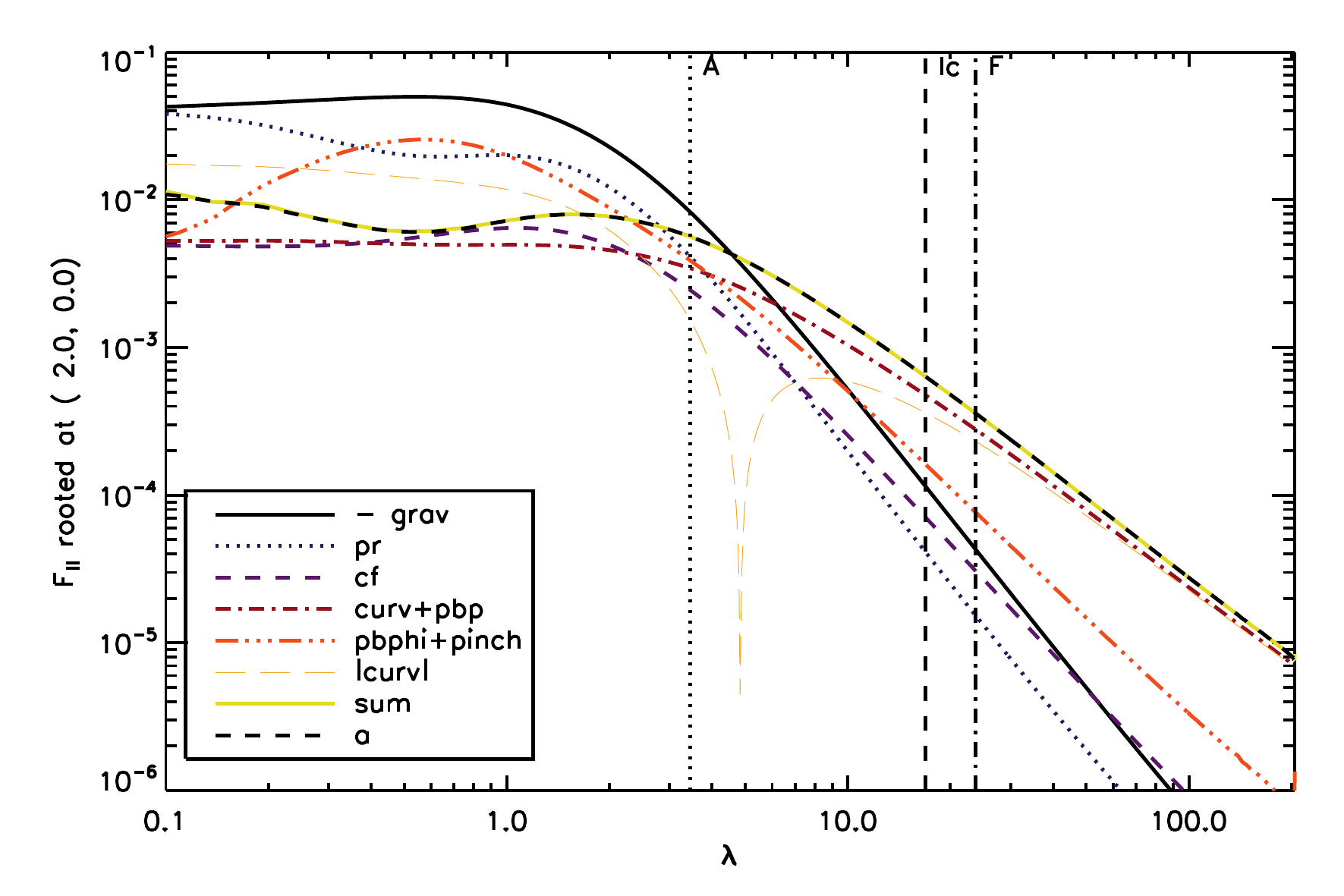}
\includegraphics[width=0.49\textwidth]{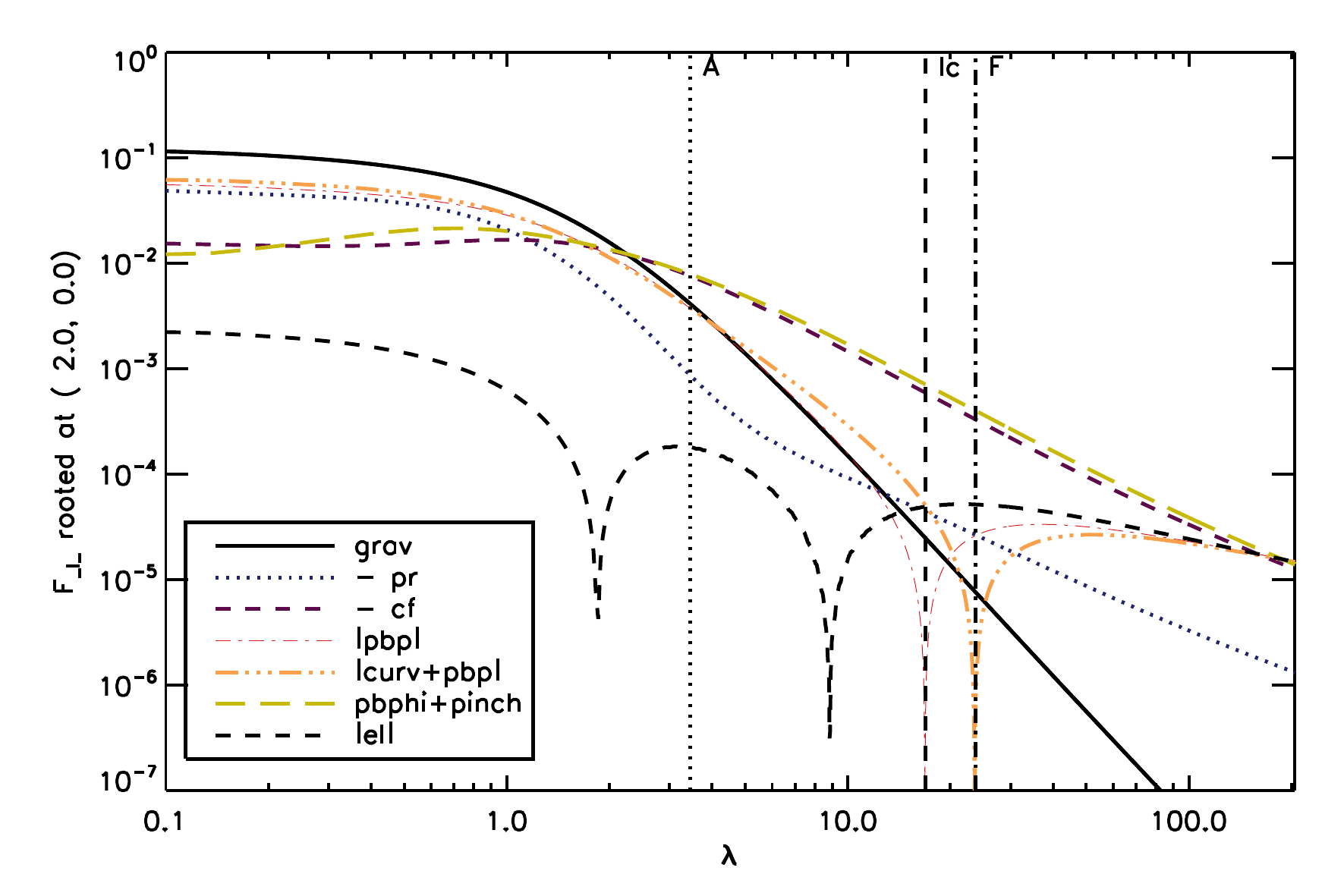}
\includegraphics[width=0.49\textwidth]{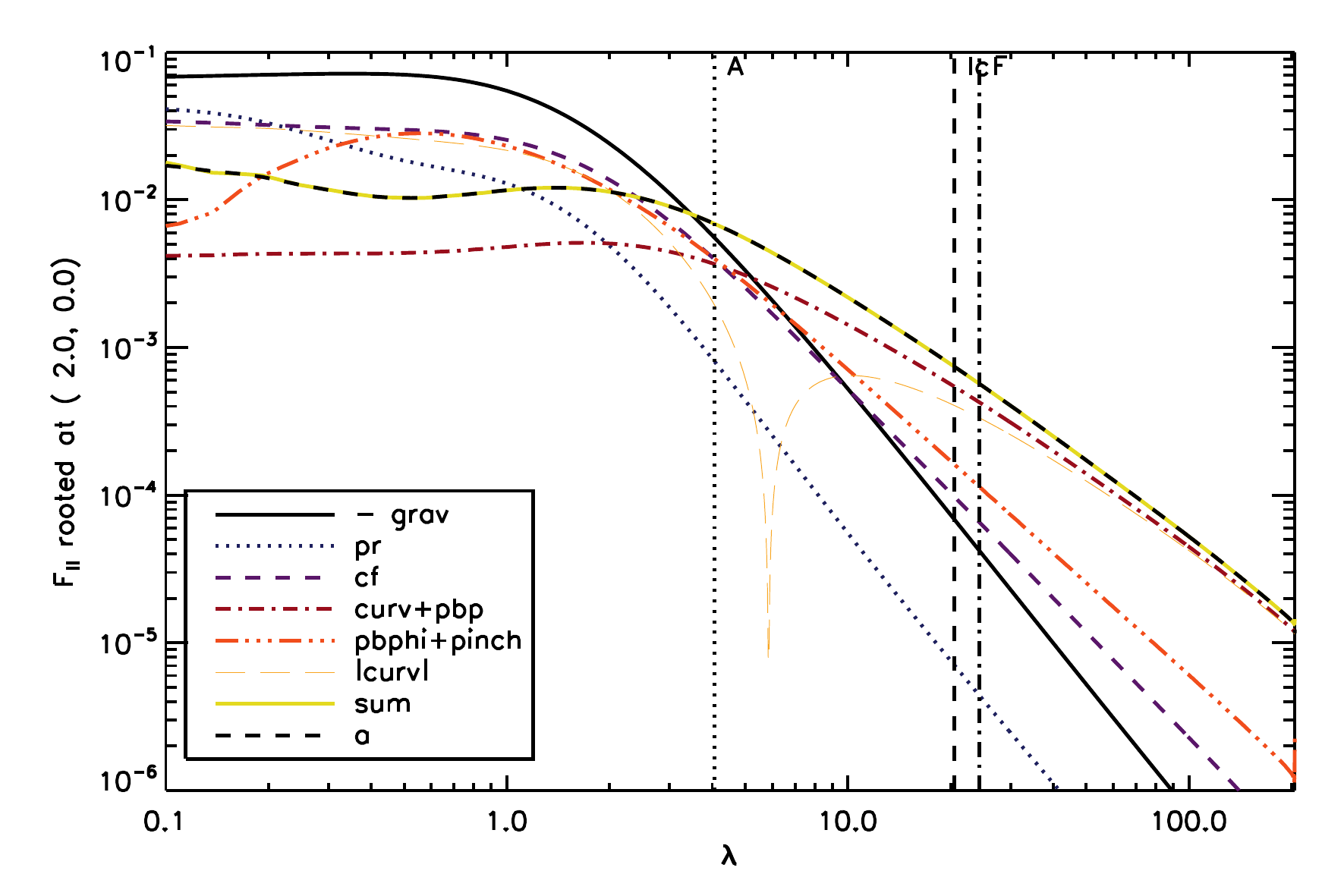}
\includegraphics[width=0.49\textwidth]{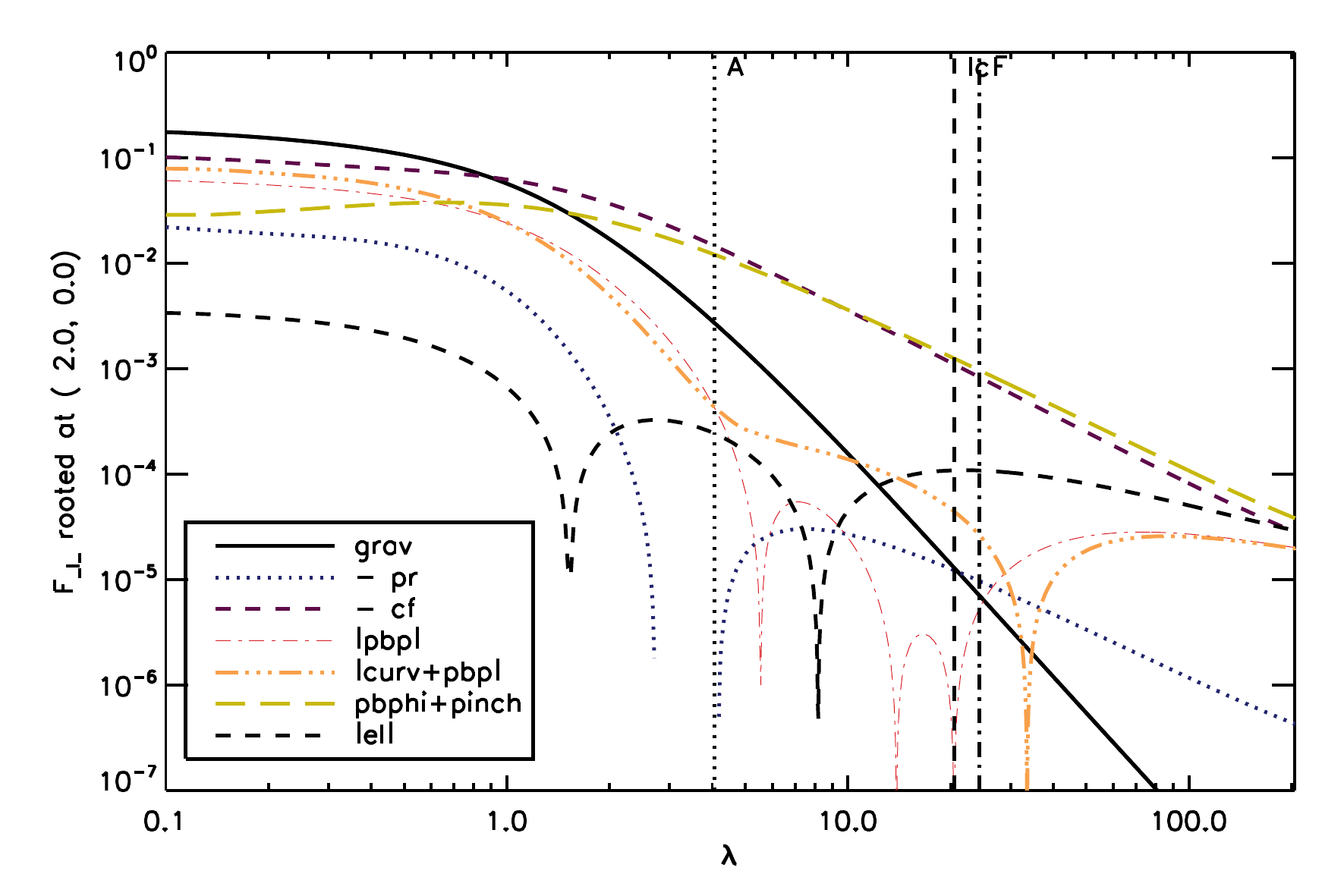}
\caption{Accelerating (left column) and collimating (right column) 
forces along the field line rooted at $r_{\rm fp}=2$ for the three
reference simulations (from top to bottom):
{\it WB01} (split monopole), 
{\it WA02} (hourglass potential field, hot case), and 
{\it WA05} (hourglass potential field, cool case).
Shown are the contributions from gravity, gas pressure gradient, 
centrifugal force, poloidal magnetic field pressure gradient, 
poloidal field tension and pressure
gradient, toroidal magnetic field pressure gradient and tension, and
forces due to the electric field.
Vertical lines indicate the critical surfaces - the Alfv\'en point along
this field line, denoted by 'A', the light cylinder radius denoted by 'lc',
and the fast magnetosonic point denoted by 'F'.
In this logarithmic representation a change of sign in the force
direction is indicated by the singularities along the graphs.
\label{fig:f-hot}}
\end{figure*}

Fig.\,\ref{fig:f-hot} compares these forces for a number of reference 
simulations (WB01, WA02, WA05)
along a field line rooted at $r_{\rm fp}=2$.
As check for consistency, we also show the gradient of 
the Mach number $a \equiv B_{p}^2/4\pi\nabla_{||}M^2$ which 
just coincides with the summation 
of the parallel forces, indicating a steady state
(see yellow solid and black dashed line).

In general, the outflow starts with sonic speed and is first launched by thermal pressure in the hot disk corona, respectively 
the centrifugal force in the colder version.
%
Until the Alfv\'en point, the Lorentz-force of the poloidal electric current 
($F_{\rm pbphi}+F_{\rm pinch}$) is the main magnetic driver.
Ultimately the poloidal tension ($F_{\rm curv}$) keeps the
acceleration up even above the fast surface.

Concerning the transverse force, we reproduce the expected sign-change 
of the curvature (tension) force (first collimating until the Alfv\'en 
surface, de-collimating beyond) and the poloidal pressure force 
(de-collimating until the light cylinder, collimating beyond).
For the cross-field balance, we observe the following three regimes: 

Just on top of the inlet, the main de-collimating forces besides poloidal magnetic pressure are thermal pressure in the hot case (WB02, WA02) and centrifugal support in the colder case (WA05).  Gravity is here the strongest force towards the origin and the situation just reflects the radial force-equilibrium we have applied for the inlet boundary.  This is the hydrodynamic regime.  

At the Alfv\'en point, the residual of the pinch- and toroidal pressure-force ($\mathbf{j_{p}}\times B_{\phi}$) is the main collimator, balanced by the centrifugal term.  Thermal pressure quickly looses importance.  
This is the magneto-hydrodynamic regime.  

In the asymptotic region beyond the light-cylinder, de-collimation by electric forces overcomes the centrifugal force and is balanced by the poloidal magnetic pressure that changes its sign at the light-cylinder (best seen in WB01).  
This is the relativistic regime.  

\begin{figure}[htbp]
\centering
\includegraphics[width=\columnwidth]{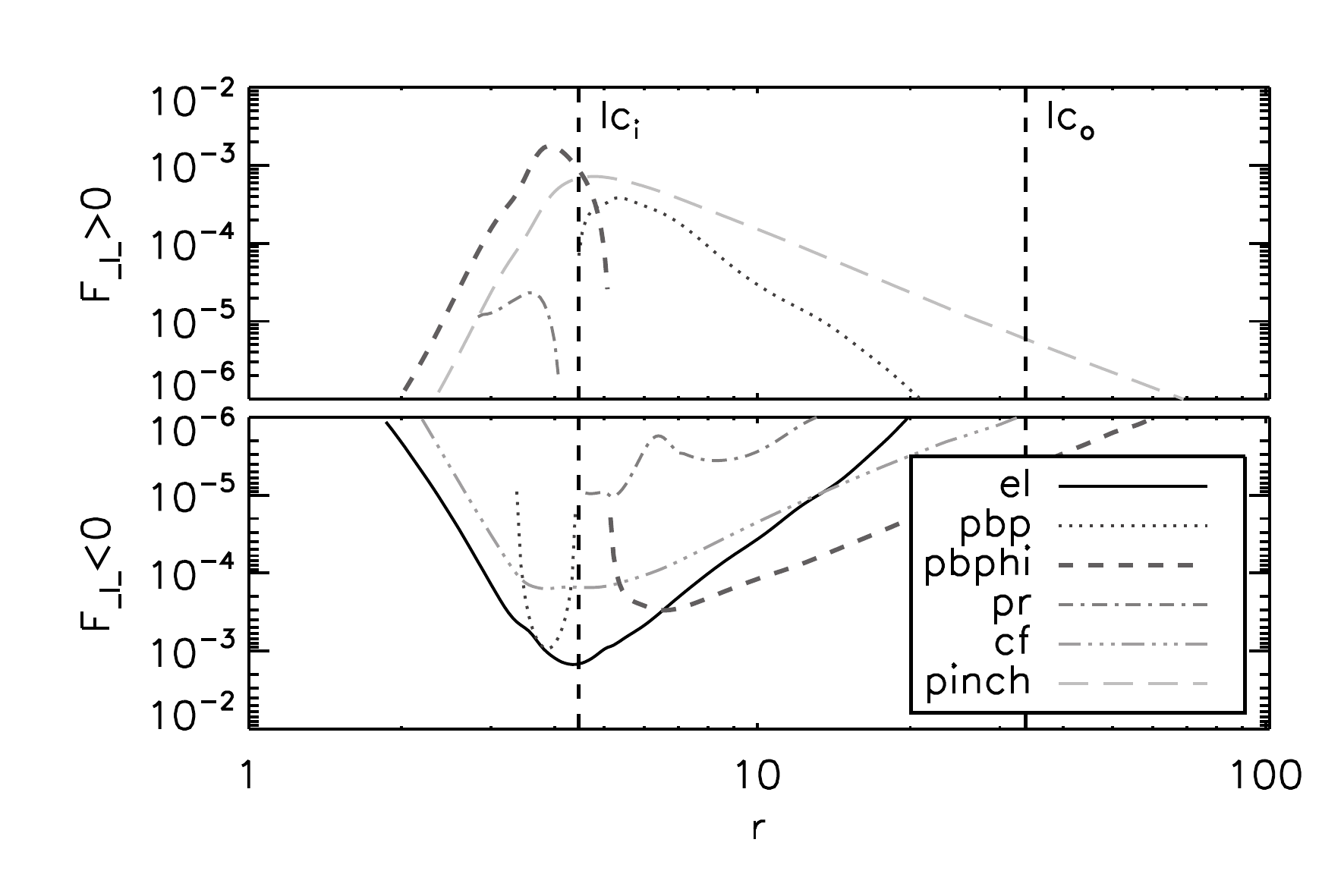}
\caption{
Trans-field force cut at $z=200$, inner ($lc_{i}$) and outer ($lc_{o}$) light-cylinder.  The differentially rotating field lines are fastest at the inner disk-radius, resulting in the inner light-cylinder - here electric de-collimation is important.  The $B_{\phi}(r)$ profile is curled up from the inner disk radius onwards and results in a magnetic pressure gradient that works in unison with the pinch force until the toroidal field surpasses its maximum. Within $r<3$, we omit the curves for thermal and poloidal pressure. These terms fluctuate around $\pm10^{-5}$ while balancing each other.  
 (run WA02)
\label{fig:force-balance}}
\end{figure}

To get a global impression on the relative importance of the individual
forces we show a radial cut throughout the asymptotic jet in figure
\ref{fig:force-balance}.
The strongest forces arise across the inner asymptotic light 
surface which separates field lines of high angular velocity from 
those in the non-rotating corona along the axis.
Here the electric de-collimation is essential.
The $B_{\phi}(r)$ profile is curled up from the inner disk radius along 
the outflow - resulting in a magnetic pressure gradient that works 
in unison with the toroidal field pinch force until at some radius the 
toroidal field surpasses its maximum and decreases (negative gradient).

The strong gradients in toroidal field and rotation induce a current sheet
and give rise to an electric charge.
The space charge $\rho_{\rm e}=(1/4)\pi\nabla\cdot \mathbf{E}$ is positive
close to the axis and changes its sign at a critical line as defined
by \cite{1969ApJ...157..869G}.

\subsubsection{Energy conversion}
In the simulations where injection is sub-magnetoslow, the energy flux is 
not a free parameter, but is consistently determined by the simulation of the disk-wind. 
It is hence of interest how the partitioning and conversion is realized.    
From the values of $\mu_{\rm max}$ given in Tab.\,\ref{tab_all} it is obvious 
that our disk-corona supports only mildly relativistic flows 
below $\Gamma=1.5$ (section \ref{sec:flconst}).  

\begin{figure*}
\centering
\includegraphics[width=0.49\textwidth]{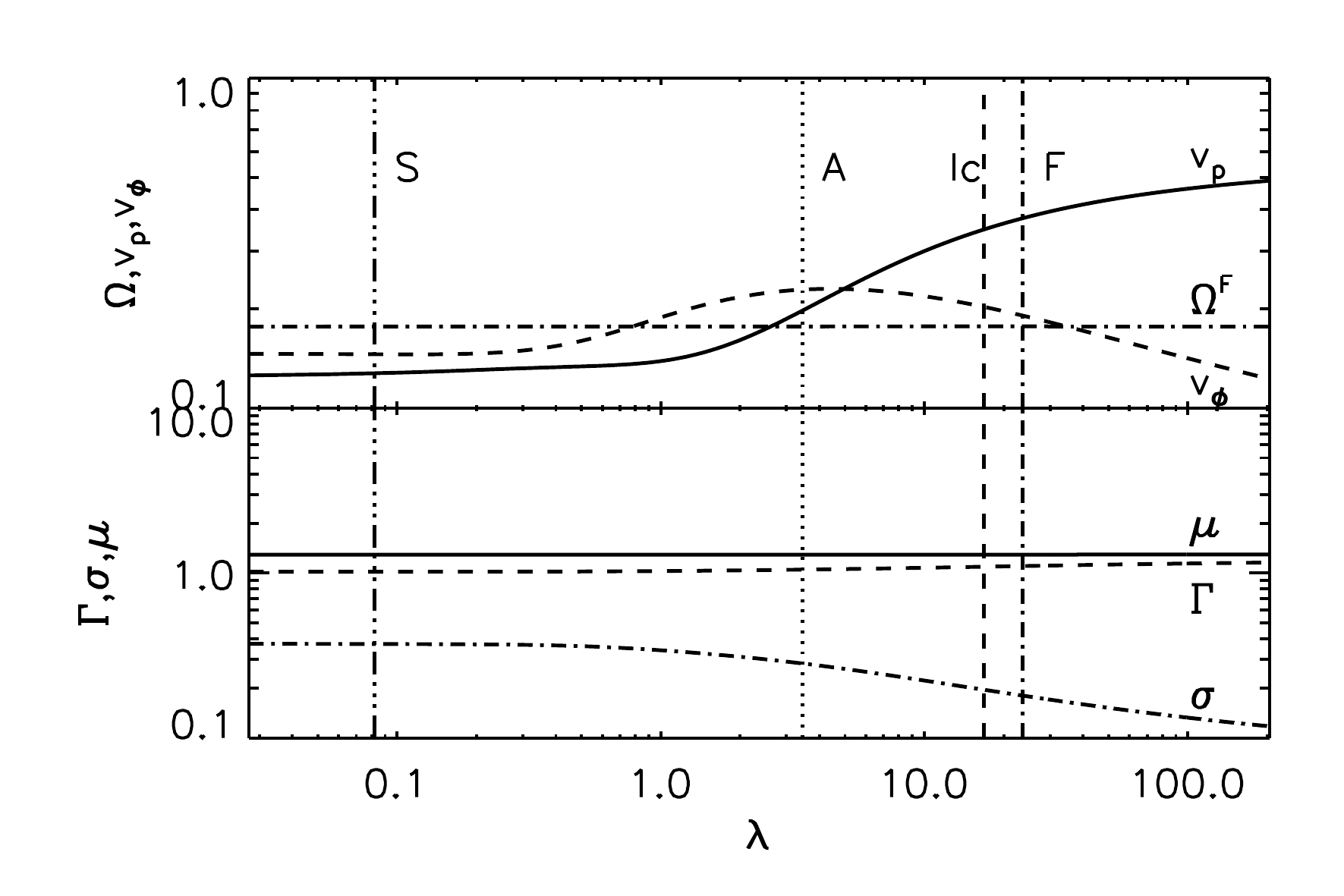}
\includegraphics[width=0.49\textwidth]{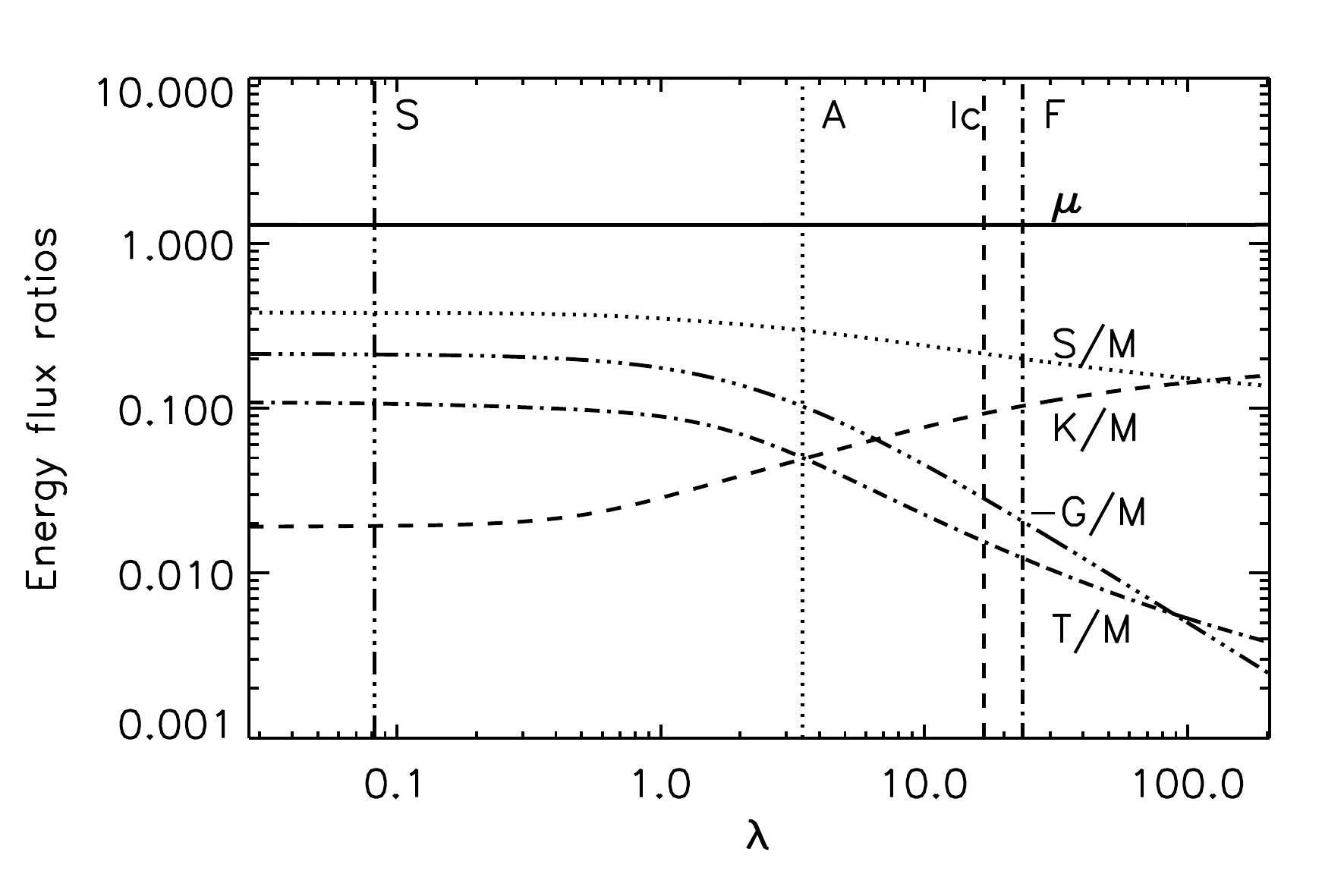}
\caption{ 
Dynamical quantities as a function of the distance along the 
field line rooted at $r_{\rm fp}=2$ for the model WA02. 
Vertical lines indicate the slow-magnetosonic, Alfv\'en, light surface and fast magnetosonic transitions (from left to right).
\textit{Left:} 
Isorotation parameter $\Omega^F$, poloidal and toroidal velocity are shown in the top panel. 
Lorentz factor $\Gamma$, energy conversion efficiency $\sigma$, 
and normalized total energy flux $\mu$ in the bottom panel.  The flow is below equipartition already at the base of the jet.  
\textit{Right:}
Complete energy flux ratios.  In the asymptotic region, thermal and gravitational fluxes are obviously negligible. 
\label{fig:vp}}
\end{figure*}

In Fig.\,\ref{fig:vp} (bottom left panel) we show the efficiency $\sigma$ of Poynting flux to kinetic 
flux conversion along the field line with $r_{\rm fp}=2$ in the fast component of the jet.  
Here, $\sigma$ is below equipartition already at the inlet and it further 
decreases as $\Gamma$ approaches $\mu$.  
In Fig.~\ref{fig:vp} (top left) the toroidal velocity shows that the flow decouples 
from co-rotation with the magnetic field at the Alfv\'en point.
Beyond the Alfv\'en point, angular momentum is then carried predominantly
by the magnetic field.  
The poloidal velocity increases from low injection value (sonic velocity) to 
$\sim0.5c$. 
Further acceleration can not be expected as the bulk of the energy is already 
in kinetic form.  
The right panel of Figure \ref{fig:vp} shows the individual energy channels compared to the
rest-mass flux for the same field line. 
At the base of the jet, the strong poloidal electric currents (a strong toroidal field)
give rise to an outflow with $\mathcal{K}<\mathcal{T}<\mathcal{-G}<\mathcal{S}<\mathcal{M}$, predominantly transporting energy via rest-mass and Poynting-flux.  

The kinetic energy flux surpasses the thermal flux at the Alfv\'en point and 
further overcomes the gravitational binding energy term shortly thereafter.  This is not surprising, since the escape-surface can be close to the Alfv\'en surface at least for the inner field lines (see also Fig. \ref{fig:stationary_flow}).

Only then, the cold limit $\mu = \Gamma(\sigma+1)$ is applicable - it is certainly 
valid in the asymptotical outflow where thermal and gravitational energy fluxes 
are negligible.


\subsection{Dependence on the launching environment}
For the simulations described up to now, we have performed in addition several 
parameter runs in order to investigate how the resulting jet dynamics
depends on the (prescribed) launching conditions - the disk corona
(see Tab.~\ref{tab_all}). 
We now focus on the impact of the plasma $\beta$ and the disk temperature 
parameter $\epsilon$.  

In general, a low $\beta$ (a stronger magnetic field) we find that the outflow tends
to collimate more, as indicated by the higher average collimation degree $\xi$ 
and a lower momentum weighted jet radius $r_{\rm jet}$.  
This in principle decreases the MHD acceleration efficiency which critically 
depends on the divergence of flux surfaces.  
It is straightforward to define the mass flux-weighted (half-) opening angle of 
outflow,
\eqon
	\theta_{\dot{M}} = \rm atan\ \xi^{-1},
\eqoff
which translates to angles of $3^{\circ}<\theta_{\dot{M}}<7^\circ$ for the 
outflows under consideration.

The impact of the magnetic field strength on the amount of mass flux is not clearly visible, 
as the two simulations with $\epsilon=1/6, 2/3$ show a different trend. 
As the $\epsilon$-parameter is simply a proxy for the disk corona density, 
it will affect the collimation in the following manner: 
A higher inflow density lowers the Alfv\'en surface towards the disk
surface which in turn broadens the current topology 
and therefore widens the flow.  This is also the trend that we observe in the indicators $\xi$ and $r_{\rm jet}$.  
Given that the injection speed calculated iteratively from the outflow
simulation approaches the slow-magnetosonic speed, 
we expect the mass flux to scale as $\dot{M}\propto \sqrt{p \rho}$.  
In fact, this is approximately realized since we have 
$\dot{M}(\epsilon=1/6)/\dot{M}(\epsilon=2/3) =\sqrt{2.5}\simeq 1.6$.  

The change of the initial split-monopole inclination $\theta$ has little effect
on the overall jet collimation angle.
In particular, we observe an opposite trend as
the wider initial field with $\theta=77^\circ$ ends up slightly more
collimated than the one with $\theta=85^\circ$.  
Clearly a wider initial field leads to a larger jet radius $r_{\rm jet}$.  
Here, we like to stress the point that for the final steady state solutions
in our simulations the initial field structure is important only insofar
as it also prescribes the poloidal magnetic field profile along the outflow
launching boundary. 
The field structure is completely changed from the initial steady structure
to a new dynamic equilibrium.
Thus it makes no sense to compare the collimation of the initial field
with the collimation of the outflow field distribution.

\par 
Having pointed out the crucial role of the 
\red
vertical energy flux from the disk surface 
\black
$\mu$ and the closely
related quantity $\sigma = \mathcal{S/(K+M)}$, 
we now study the two most promising handles in increasing $\mu$.  
That is (i) a decrease in mass flux $\mathcal{M}$ and 
        (ii) an increase in Poynting-flux $\mathcal{S}$.  
We first focus on (i) and describe (ii) thereafter.  

\subsubsection{Towards low mass loading}\label{sec:light-winds}
\red
A way to obtain high-speed jets seems to be a lower mass
load injected into a similarly-strong magnetic flux.
According to the well-known Michel-scaling \citep{1969ApJ...158..727M}, 
the asymptotic outflow velocity depends on the mass flux 
$u_{\infty}\propto \dot{M}^{-1/3}$.  
We investigate this interrelation running another set of simulations 
where we change the mass flux by prescribing a low injection velocity
(instead of lowering the density of the injected gas,
 see Tab.~\ref{tab:extended1})

\begin{deluxetable*}{ccccccl||crrrrr}
\tablecaption{Effect of mass-loading
\label{tab:extended1}}
\tabletypesize{\scriptsize}
\tablewidth{\textwidth}
\tablehead{
 \colhead{ID}                      & 
 \colhead{Top}               &
 \colhead{$\beta$}                     & 
 \colhead{$\epsilon$}       & 
 \colhead{$v_{\rm inj} $}        & 
 \colhead{$\eta$}         &
 \colhead{Remarks}      &
 \colhead{$\Gamma_{\rm max}$} &
 \colhead{$\mu_{\rm max}$} &
  \colhead{$\xi$} &
 \colhead{$v_{p,\rm max}$} &
 \colhead{$r_{\rm jet}$} &
 \colhead{$\dot{M}$}
 }
\startdata
I001 & A & 0.2  & 2/3  &  0.01  &  var.  & - &  
1.66&         2.01 &        25.61 &       0.75 & 
       21.11 &        10.63
\\

I01 & A & 0.2  & 2/3  &  0.1  &  var.  & - &  
       1.52&         1.76 &        25.58 &       0.71 & 
       20.49 &        13.78
\\

I02 & A & 0.2  & 2/3  &  0.2  &  var.  & - &  
1.38&         1.55 &        26.02 &       0.65 & 
       19.34 &        17.53
\\

I04 & A & 0.2  & 2/3  &  0.4  &  var.  & - &  
1.31&         1.45 &        25.30 &       0.60 &        17.07 &        25.84
\\
I08 & A & 0.2  & 2/3  &  0.8  &  var.  & - &  
1.16&         1.24 &        40.05 &       0.47 &        14.74 &        51.30
\\
I12 & A & 0.2  & 2/3  &  1.2  &  var.  & - &  
       1.21&         1.21 &        33.77 &       0.50 &        14.10 &        81.88
\enddata
\tablecomments{As in table \ref{tab_all}, but for the extended parameter study with a given $v_{\rm inj}$.}
\end{deluxetable*}

\begin{figure}[htbp]
\centering
\includegraphics[width=\columnwidth]{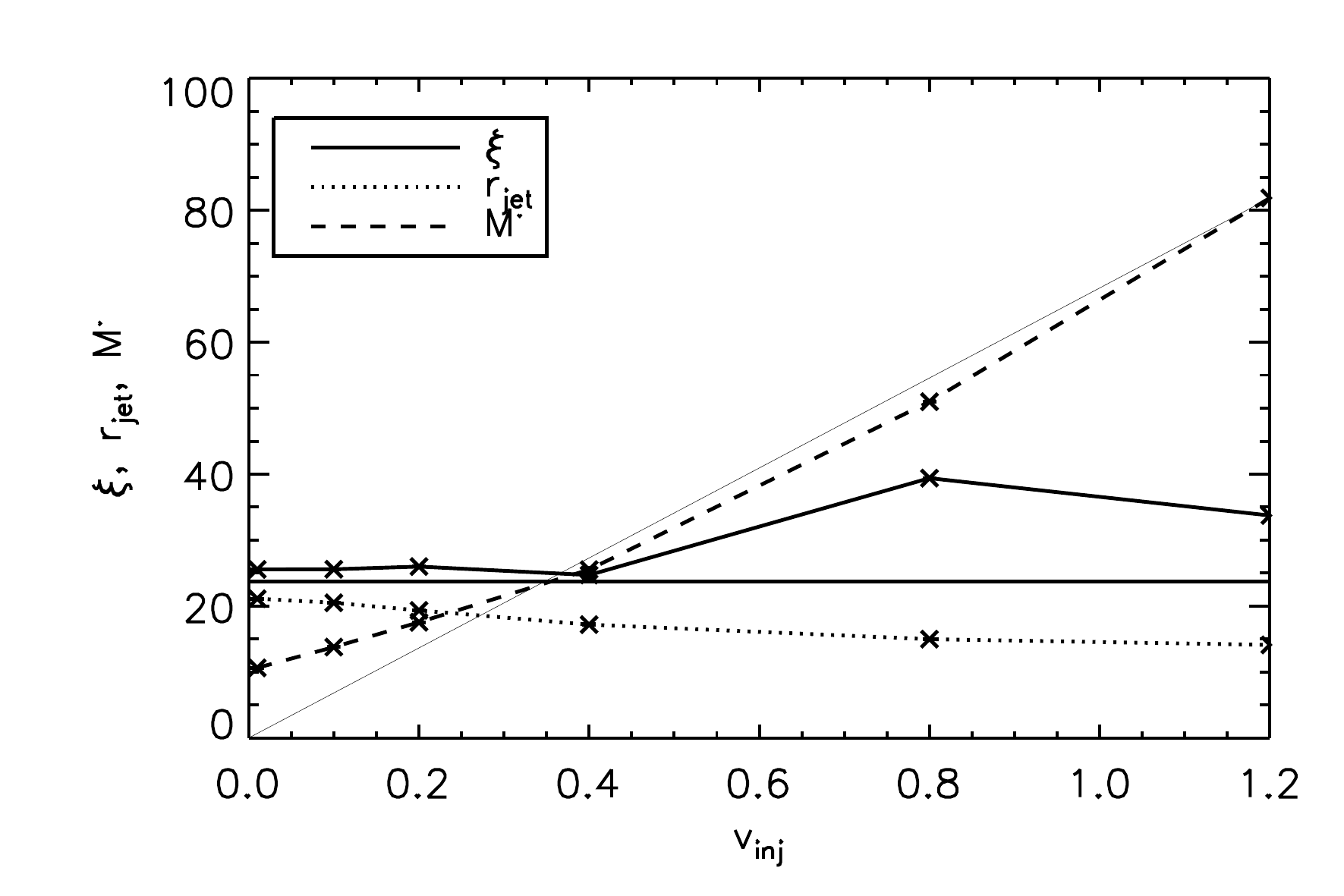}
\caption{Jet collimation and dynamics against injection speed parameter $v_{\rm inj}$, The horizontal line indicates the mass flux when $v_{\rm inj}$ is not specified (WA01).  
For $v_{\rm inj}<0.4$, the mass flux approaches an obviously unphysical offset value.
At higher $v_{\rm inj}$, the mass flux follows the expected linear dependence on the injection parameter (thin solid line).  
\label{fig:diag-comp}}
\end{figure}

However, we observe that for low injection speeds $v_{z}<0.4\ v_{\phi}$ 
the resulting (numerical) mass flux does not follow the expected linear 
relation $\dot{M} \propto v_{\rm inj}$.
Instead, for lower and lower injection speed it approaches some
seemingly unphysical offset value (Fig.~\ref{fig:diag-comp}).
This value is unphysical in the sense that it departs from the 
prescribed value at the boundary condition 
$d\dot{M}_{\rm inj}/dr = 2\pi r \rho_{\rm inj} v_{\rm inj}$.
This difficulty arises because the injected flow is sub-magnetosonic
and thus overdetermined by simultaneously assigning a profile in $\rho$ and $v_{z}$.

Although this is in principle problematic, it is not necessarely fatal 
for the investigation of low mass flux outflows.
The mass flux is governed by the magneto-slow point and is thus re-arranged
along the flow.
Indeed we find that above the magneto-slow surface the field line constants 
are very well conserved. 
This indicates that the flow dynamics transcends at the magneto-slow surface 
from a seemingly unphysical state into a MHD flow that satisfies the critical 
conditions at the magnetosonic surfaces.

This technique of self-adjusting the sub-slow mass inflow, however, turned out
to be limited towards lower mass fluxes.
The resulting mass fluxes are unfortunately, still too high and do not allow 
substantially higher magnetisation
(e.g. $\mu_{\rm max}=2.01$ for simulation I001 compared to $\mu_{\rm max}=1.33$ in run WA01).  
Figure \ref{fig:diag-comp} shows the trends concerning collimation $\xi$, jet radius $r_{\rm jet}$,
and integral mass flux $\dot{M}$ in the asymptotic flow. 
Increasing the mass flux enhances collimation while $\mu_{\rm max}$ decreases accordingly.  
For high injection speed, we observe that the wind originating from the very inner disk evolves
into a thin ballistic flow layer that has little in common with the jets we are interested in.  
\black
\subsubsection{Poynting dominated flows}\label{sec:extended}
Given the limitations mentioned above, we prescribe {\it a priori} 
the limiting energy flux parameter $\mu$ by constraining both the mass flux and 
the Poynting flux along the injection boundary - with the hope of thus providing 
a sufficiently energetic disk wind. 

To achieve this, we adopted a fixed-in-time toroidal magnetic field distribution,
$B_{\phi}\propto -\eta/r$ which necessarily changes 
$\Omega^F(r)$\footnote{
In case of a central black hole causality requires that $r\Omega^F(r)< 0.6$ 
which corresponds to the ISCO velocity for the Schwarzschild case with $r_{ISCO}=1$ 
in our scaling.}.
The toroidal field distribution following a $1/r$ profile corresponds to $j_{z}=0$ 
and has a profound physical motivation as it anticipates the radial currents expected
in the disk-corona.  

Table \ref{tab:extended2} summarizes the simulation runs performed within this setup. 
These simulations have a considerably stronger toroidal magnetic field at the injection
point.
We apply up to $B_{\phi}\sim 4 B_{p}$.
\begin{deluxetable*}{ccccccl||crrrrr}
\tablecaption{Poynting dominated flows
\label{tab:extended2}}
\tabletypesize{\scriptsize}
\tablewidth{\textwidth}
\tablehead{
 \colhead{ID}                      & 
 \colhead{Top}               &
 \colhead{$\beta$}                     & 
 \colhead{$\epsilon$}       & 
 \colhead{$v_{\rm inj} $}        & 
 \colhead{$\eta$}         &
 \colhead{Remarks}      &
 \colhead{$\Gamma_{\rm max}$} &
 \colhead{$\mu_{\rm max}$} &
  \colhead{$\xi$} &
 \colhead{$v_{p,\rm max}$} &
 \colhead{$r_{\rm jet}$} &
 \colhead{$\dot{M}$}
 }
\startdata

M02 & A & 0.2  & 2/3  &  0.1  &  2  & - &
       2.18&         3.29 &        27.55 &       0.86 &        23.71 &        22.48
\\

M04 & A & 0.2  & 2/3  &  0.1  &  4  & - & 
       3.51&         7.46 &        8.96 &       0.94 &        29.00 &        48.85
\\

M08 & A & 0.2  & 2/3  &  0.1  &  8  & - &
       6.11&         25.61 &        6.8377 &       0.97 &        35.89 &        80.40
\enddata
\tablecomments{As in table \ref{tab_all}, but for the extended parameter study with a given $v_{\rm inj}$ and $\eta$.}
\end{deluxetable*}

An exemplary process enhancing large-scale toroidal fields in the disk corona 
could be the MRI-driven dynamo under current investigation by many authors 
\citep[e.g.][]{2000ApJ...534..398M, 2003A&A...398..825V}.
%
The jet eventually evolving from these disks is not propelled by the
\citet{1982MNRAS.199..883B} mechanism, but driven by the toroidal magnetic 
pressure \citep{1996LNP...471...74C, 2003ApJ...599.1238D, 2004ApJ...605..307K}. 
These so called Tower jets have initially been proposed by \cite{1996MNRAS.279..389L} and directly extract Poynting-flux from the dist rather than first converting rotational energy into the twisted magnetosphere that is present at the Alfv\'en surface.\footnote{See \cite{2007Ap&SS.307...11K} for a review. 
Note also that these jets have successfully been reproduced by \cite{2005MNRAS.361...97L} in laboratory experiments with purely radial
current distributions at the base.}

\begin{figure}[htbp]
\centering
\includegraphics[width=\columnwidth]{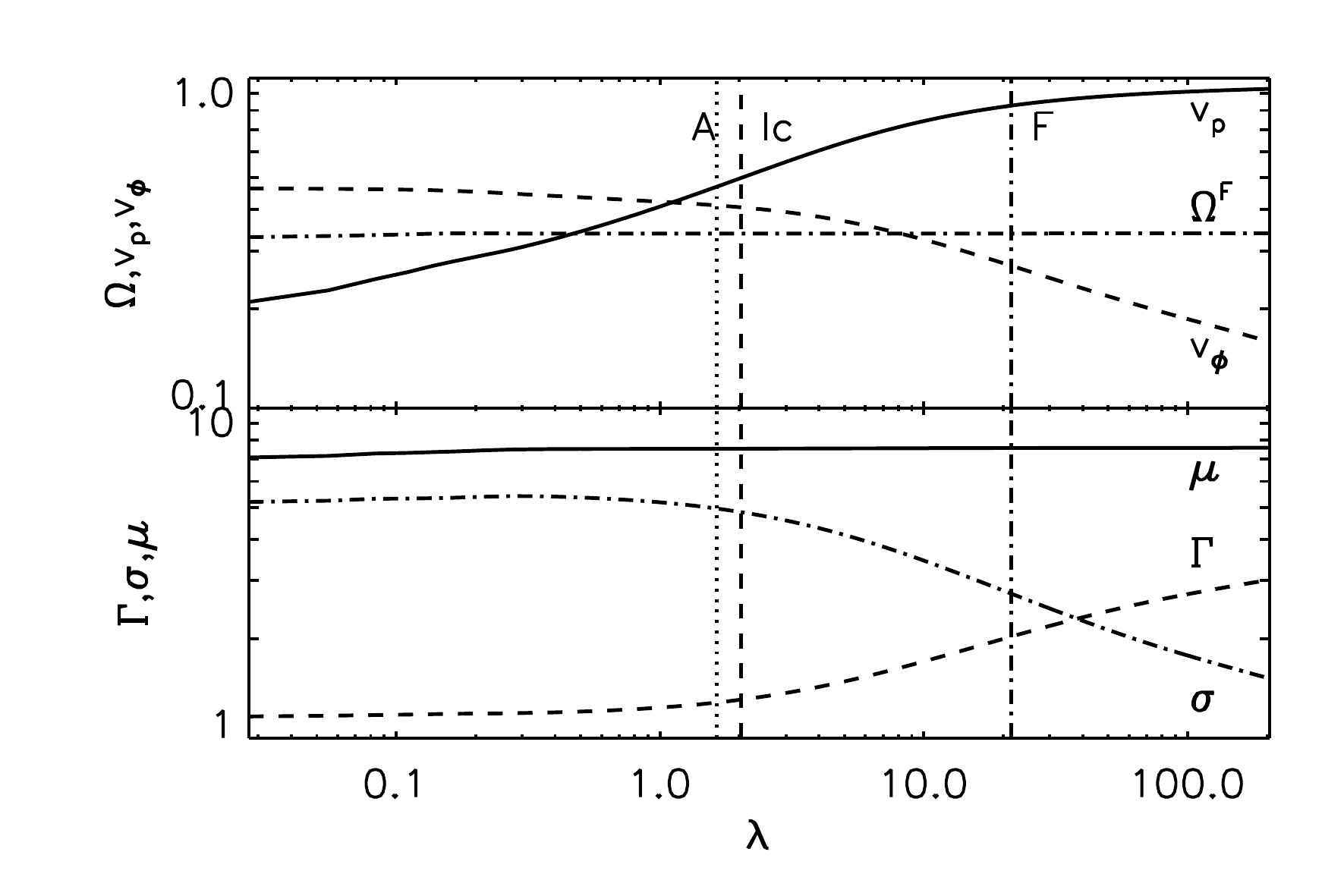}
\caption{As in Fig.~\ref{fig:vp}, however calculated for simulation run M04.  The injected flow is strongly magnetized and remains Poynting-dominated when leaving the computational domain. 
\label{fig:vp-ext2}}
\end{figure}

The simulations described here are of a mixed type, since they combine large-scale open field lines with a toroidal field emerging from a radial current.  
For an example simulation of this type, we show the conversion of energy in 
Fig.~ \ref{fig:vp-ext2} similar to Fig.~\ref{fig:vp} for the low-energy case.  
By design, the injected Poynting flux surpasses the rest mass flux with $\sigma\simeq5$ 
within the fast outflow component.  

The outflow, which is initially Poynting flux dominated, does not reach 
equipartition at the fast magnetosonic surface $r=r_{\rm F}$, where we merely find $\Gamma\sim\mu^{1/3}$ following \cite{1969ApJ...158..727M,1998MNRAS.299..341B}.  
In the asymptotic region $r\gg r_{F}$, the length scales for additional flow acceleration 
and collimation would increase exponentially with $\Gamma\propto (\mu \ln r)^{1/3}$ 
(e.g. \cite{1994PASJ...46..123T}), which is clearly beyond the reach of our numerical 
method.  


Our simulations indicate that, given sufficient Poynting flux, bulk Lorentz factors derived
from AGN jet observations can be obtained within several hundred Schwarzschild radii, 
$\Gamma \simeq 6$ for model M08. 
Since acceleration has proven to be most effective around the Alfv\'en point 
which is expected to be very close to the central object ($r_{\rm A}<r_{\rm lc}$), 
we expect this conclusion to remain valid also for higher $\mu$ and thus higher 
terminal $\Gamma$ flows.  
However, we note that when increasing $\sigma$, the energy conversion efficiency decreases, 
a situation commonly denoted as $\sigma$-problem in pulsar winds 
\citep{1974MNRAS.167....1R, 1984ApJ...283..694K}.
Several authors have recently addressed this issue with partly controversial results
\citep{2009MNRAS.394.1182K, 2009ApJ...699.1789T, 2009ApJ...698.1570L},
so that after the successful acceleration towards relativistic speeds, Poynting-flux could still remain and one is tempted to ask: Is there a $\sigma$-problem for AGN-jets? 
The simulations presented here are not fit to answer this question satisfactory.
However, we certainly know that the mildly relativistic disk winds presented earlier
do not suffer from this, as they are launched already in sub-equipartition.  

\section{Summary}\label{sec:conclusions}
We have presented ideal MHD simulations of the formation of special relativistic disk 
winds using the PLUTO 3.0 code.  
On the technical side, the key points are: \\
\textit{i)} 
The inclusion of (Newtonian) gravity allows us to specify an astrophysically sensible boundary
condition of a hydrodynamically stable disk corona.
We can thus consistently follow the acceleration from initially sub-escape velocity winds.  \\
\textit{ii)} 
Much dedication has been put in the development and testing of a novel realization for the
outflow boundary that enables us to simulate for hundreds of inner disk rotations while 
minimizing spurious collimation due to artificial boundary currents.  
Our detailed study of jet collimation is possible only through this effort.  

As a general result we obtain well collimated jets with a mass flux weighted half-opening
angle of $3 - 7^\circ$ and mildly relativistic velocities depending on the launching
conditions for the outflow.
The flow collimation happens mainly in the classical (non-relativistic) regime before the 
light surface.   
A major result of our simulations is that we - for the first time - self-consistently 
calculated the shape of that light surface.
The light surface determines the "relativistic" charater of the flow. Material which
traverses the light surface experiences the full relativistic effects.  

We can identify three dynamically distinct regions in terms of flow collimation. \\
\textit{i)} 
In the hydrodynamic regime upstream of the Alfv\'en surface, 
gravity balances thermal and magnetic pressure, respectively the centrifugal force in the colder case. \\
\textit{ii)} 
In the magneto-hydrodynamic regime following the the Alfv\'en surface downstream,
the residuals of magnetic pinch and the toroidal magnetic pressure gradient balances
the centrifugal force.  \\
\textit{iii)} 
In the relativistic regime located downstrem of the light surface, the poloidal magnetic 
pressure gradients now impose a collimating force against electric field de-collimation.
Electric forces ultimately overcome the classical magneto-centrifugal contribution.  

A steep rotation profile of the field line as given by a Keplerian disk
results in a light surface geometry which steepens for large radii.
Depending on the magnetic field profile, the light surface may even
collimate along the flow for large radii.
In such a case the relativistic core inside the light surface is naturally confined 
by a non-relativistic wind.
The ability of both the relativistic jets and the non-relativistic disk winds to collimate may provide confining agents for an axial ultra-relativistic funnel 
which could probably launched by the Blandford-Znajek process.  

The relatively slow winds found to arrise at large distances of around 100 Schwarzschild radii may be observed as X-ray absorption winds in radio-quiet AGN.  

In the case of Blandford-Payne disk winds ($B_{\phi}=0$ initially), the 
outflow is kinetic energy-dominated with the ratio of electromagnetic energy flux 
to kinetic energy flux $\sigma<1$ already at the jet base and with this ratio 
further de-creasing downstream the outflow.  
These disk winds start out at sonic speed and reach only mildly relativistic speeds 
up to Lorentz factors $\Gamma < 1.5$.
We have also investigated cases where the jet Poynting flux is increased by
directly injecting an additional toroidal magnetic field from the disk boundary.
In this case we achieve terminal Lorentz-factors up to $\Gamma_{\rm max}\simeq 6$
while the super fast-magnetosonic jet remains Poynting-flux dominated.  

\begin{acknowledgements}
We thank Andrea Mignone and the PLUTO team for the possibility to use the
PLUTO code and for his support with the relativistic module.  
Numerical simulations were performed on the PIA cluster of the Max Planck Institute for Astronomy (Heidelberg) located at the Rechen-Zentrum in Garching.
O.P. likes to thank Bhargav Vaidya for his commitment in numerous discussions.  
We are pleased to acknowledge clarifying conversations with Max Camenzind.
\end{acknowledgements}

\clearpage
\appendix
\section{Zero current boundary}\label{sec:zerocurrent}
One of the major goals of this paper is to investigate the collimation
behavior of relativistics outflows.
It is therefore essential to exclude any numerical artifacts leading
to a spurious flow collimation.
We find that the standard zero-gradient outflow boundary conditions
may lead to an un-physical Lorentz force in radial direction implying
such spurious collimation (or de-collimation).

Thus, we put substantial effort in implementing and testing an enhanced 
outflow boundary condition to the code.

To get a handle on the Lorentz-force $j_{\phi}\times B_{p}$, one has to address the toroidal electric currents at the grid boundary.  
In principle there are (at least) two options.
One is the possibility to copy the toroidal electric current across the
boundary.
While this approach should minimize spurious collimation efficiently, we 
observed that the overall stability of the simulation was decreased.
Thus we decide to use the following zero-toroidal current outflow boundaries
in our simulations.

\begin{figure}[htbp]
\centering
\includegraphics[width=10cm]{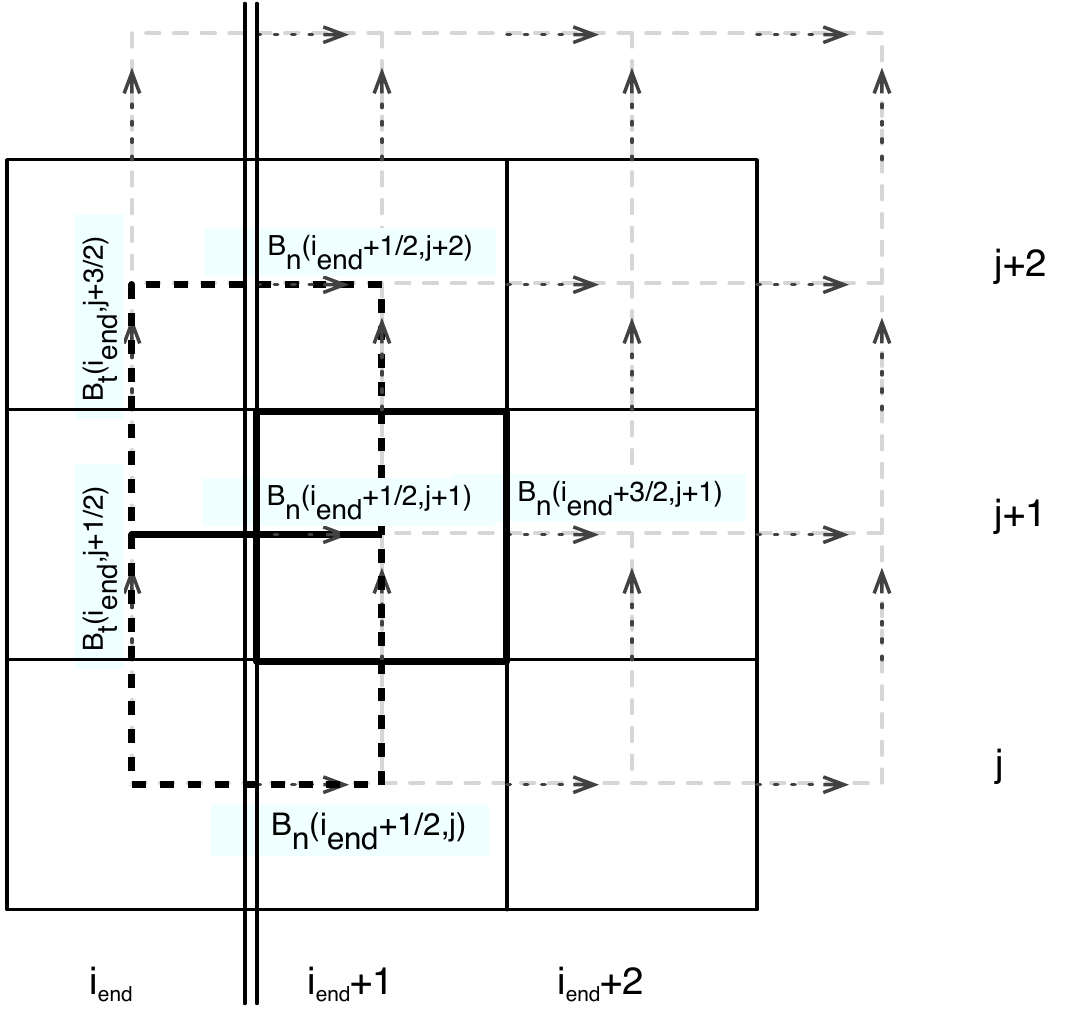}
\caption{Construction of the 
$\mathbf{\nabla \times B_{p} = 0}$ and $\nabla\cdot \mathbf{B} = 0$ 
boundary condition. 
Shown is the last grid slab of the domain $(i_{\rm end},j)$ and a ghost zone 
of two elements.   
\label{fig:zero-current}}
\end{figure}

In this case we take advantage of the staggered grid by enforcing zero 
toroidal currents while simultaneously satisfying the solenoidal condition 
$\nabla\cdot \mathbf{B} = 0$. In the following our procedure is described in 
detail.
We consider computational grid cells $(i_{\rm end},j)$, adjunct to the domain
boundary at $(i_{\rm end}+1,j)$, as illustrated in Fig.~\ref{fig:zero-current}.  
The magnetic field components of the domain, 
$B_{\rm t}(i_{\rm end},j+1/2)$, $B_{\rm n}(i_{\rm end}+1/2,j)$, 
$B_{\rm n}(i_{\rm end}+1/2,j+1)$, together with the transverse field 
component $B_{\rm t}(i_{\rm end}+1,j+1/2)$ of the first ghost zone,
constitute a toroidal corner-centered electric current 
$I_{\phi}(i_{\rm end}+1/2,j+1/2)$.  
Utilizing Stokes theorem, 
$I_{\phi} = \int\! d\mathbf{S\cdot\nabla\times B_{p}} = \oint\! d\mathbf{l\cdot B_{p}}$, 
we then solve for the unknown field component 
$B_{\rm t}(i_{\rm end}+1,j+1/2)$ under the constraint that $I_{\phi}=0$,
\begin{equation}
B_{\rm t}|_{i_{\rm end}+1,j+1/2} = 
B_{\rm t}|_{i_{\rm end},j+1/2}+ 
\frac{\Delta r}{\Delta z}\left[B_{\rm n}|_{i_{\rm end}+1/2,j+1}-
       B_{\rm n}|_{i_{\rm end}+1/2,j}\right]\label{eq:bt}
\end{equation}
where we have assumed an equally spaced grid for clarity of the argument.  
Once $B_{\rm t}(i_{\rm end}+1,j+1/2)$ is known for all $j$, the next layer of 
normal field components $B_{\rm n}(i_{\rm end}+3/2,j)$ can be inferred from  
the $\nabla\cdot \mathbf{B} = 0$ constraint in its integral form, 
\begin{equation}
B_{\rm n}|_{i_{end}+3/2,j+1} = 
\frac{\Delta S_{\rm n}B_{\rm n}|_{i_{\rm end}+1/2,j+1}
+\left( \Delta S_{\rm t} B_{\rm t}|_{i_{\rm end}+1,j+1/2}- 
\Delta S_{\rm t} B_{\rm t}|_{i_{\rm end}+1,j+3/2}\right)}{\Delta S_{\rm n}|_{i_{\rm end}+3/2,j+1}}.
\end{equation}
For the next grid layer, the transverse field components can again be found 
applying Eq.~\ref{eq:bt}, and the process is repeated for the each layer.  

Some words of caution. 
We find that the current-free magnetic field boundary condition can only be realized 
when the grid cell aspect ratio $\Delta z/\Delta r$ is not too large.  
An aspect ratio of e.g. $12/1$ resulted in errors of $100\%$ in $B_{\rm n}$ 
at the most critical areas close to the symmetry axis leading to an overall 
unstable flow evolution. 
We find that as a rule of thumb, an aspect ratio of $3/1$ should not be exceeded. 
We also emphasize that it is essential to treat the grid corners consistently.
This is because field components in the corner,
$B_{\rm t}(i_{\rm end}+1,j_{\rm end}+1/2)$, and $B_{\rm n}(i_{\rm end}+1/2,j_{\rm end}+1)$,
are interrelated which would lead to an ambiguity.
In order to avoid this ambiguity, we decided to extrapolate the values in
question which does provide the information that is missing otherwise.  

\begin{figure}[htbp]
\centering
\includegraphics[width=0.25\textwidth]{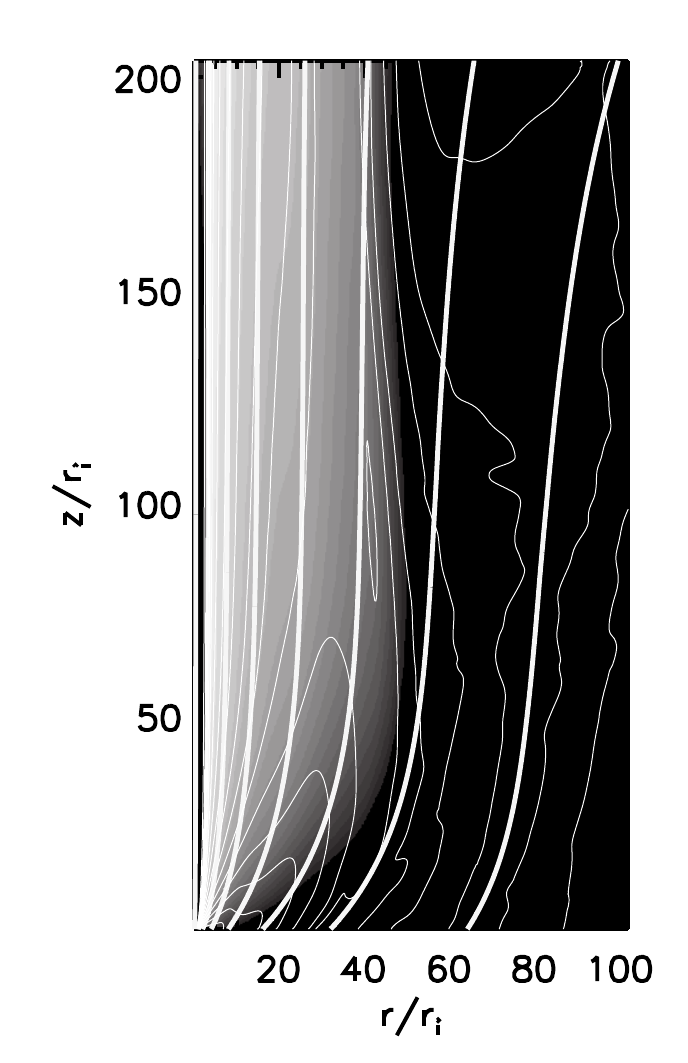}
\includegraphics[width=0.25\textwidth]{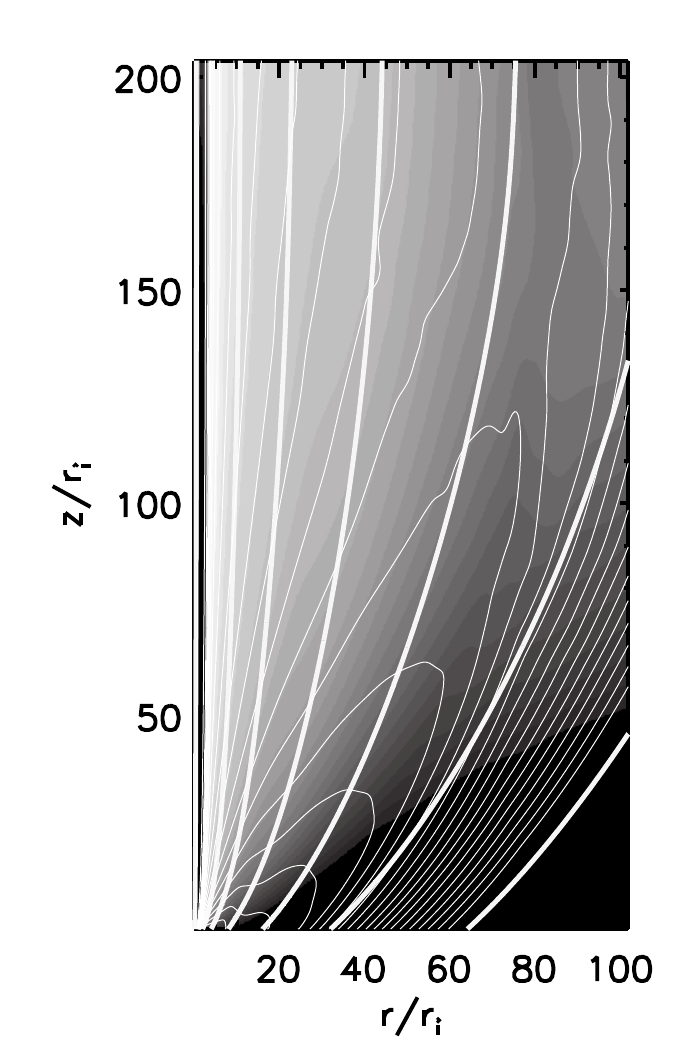}
\includegraphics[width=0.25\textwidth]{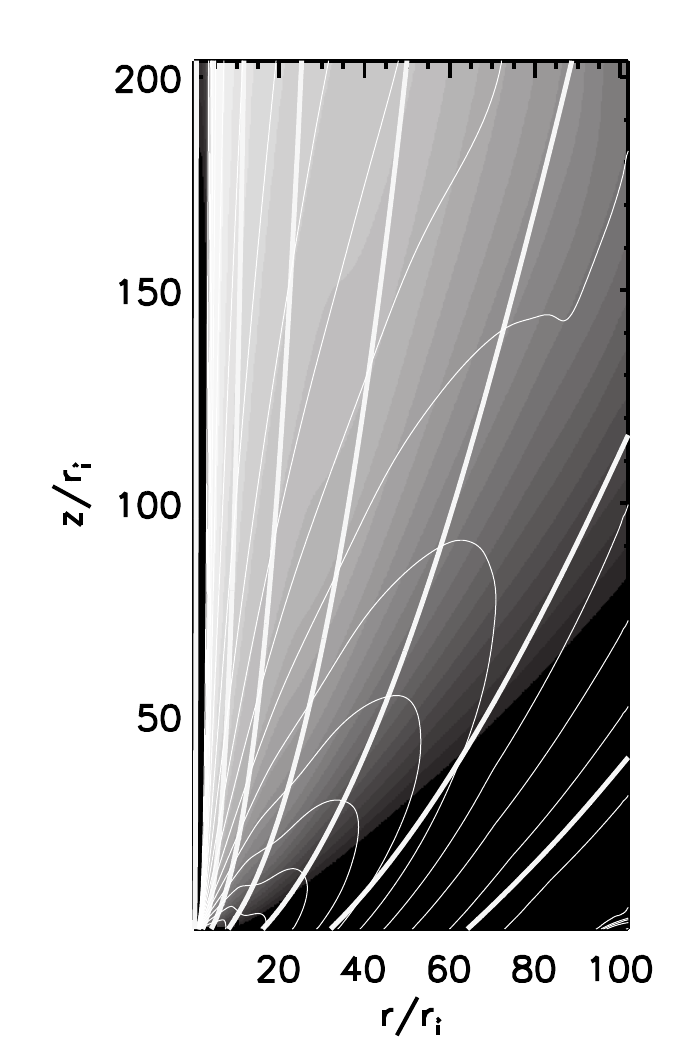}
\caption{Comparison of simulations applying a variation of outflow boundary 
conditions for the magnetic fields at the time of 100 inner disk rotations. 
The parameters are equal to those in simulation WA04. 
The grayscale indicates the Lorentz factor $log(\Gamma-1)$ as in figure \ref{fig:simul}, 
the poloidal magnetic field (the poloidal electric current) is shown in 
thick (thin) white contours. 
Standard zero gradient, 
zero second derivative,
zero current boundary condition, respectively (from left to right).
\label{fig:current-comp}}
\end{figure}

We demonstrate quality of our approach by showing results of simulations which 
{\em do not} apply the zero current but the zero-gradient or the zero second derivative outflow condition 
with otherwise the same flow parameters as in simulation
WA04 (Fig.~\ref{fig:current-comp}). 
As it can be seen, for zero-gradient boundary conditions, the effect of 
collimation by artificial currents is so strong that no steady-state can be
reached and the flow is continously squeezed towards the axis.  Also in zero second derivative, we observe an artificial alignment with the grid geometry.  

\begin{figure}[htbp]
\centering
\includegraphics[width=0.4\textwidth]{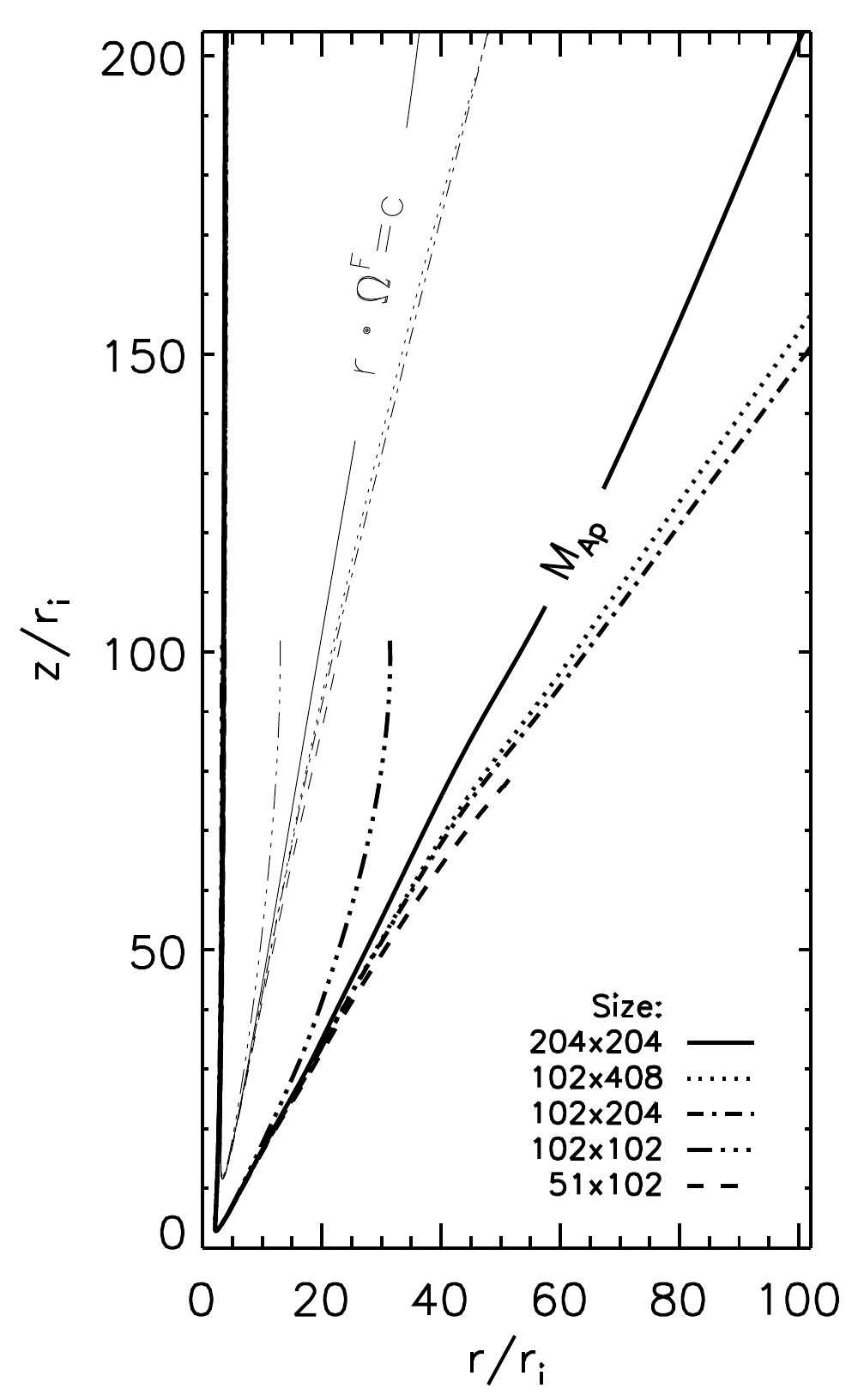}
\includegraphics[width=0.4\textwidth]{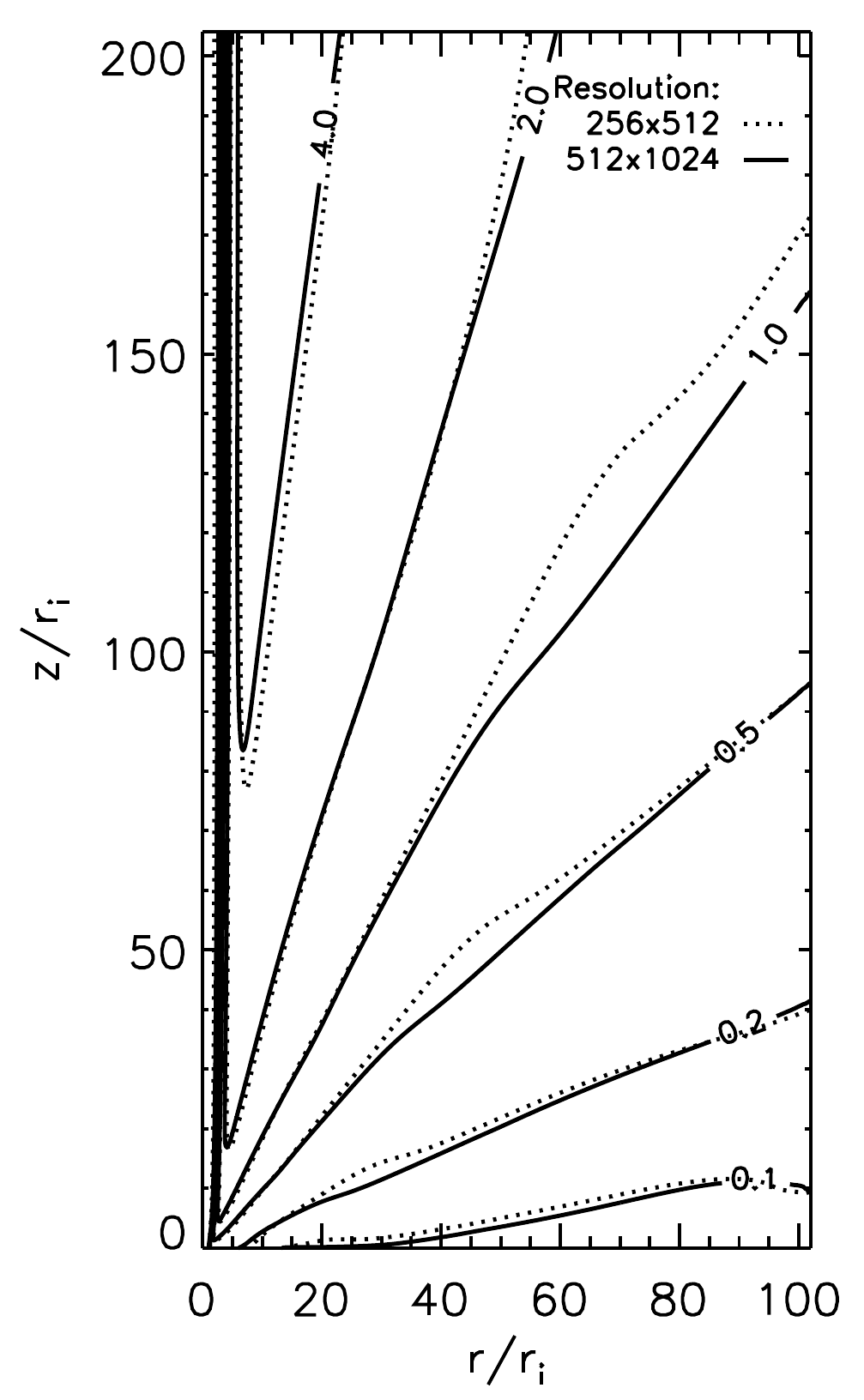}
\caption{
Comparison of different grid realisations with zero current outflow boundaries after 200 inner disk rotations.  Parameters as in simulation WA04.   
\textit{Left:} 
Various geometries and sizes.  Shown is the critical Alfv\'en surface and the light cylinder.  
\textit{Right:}
Convergence test with two grid resolutions.  Shown are contours of the Alfv\'en mach number $M$ for $256 \times 512$ and for $512 \times 1024$ grid elements.  
\label{fig:box-comp}}
\end{figure}

We check the geometry dependence of the zero current outflow boundary by several realizations of the fiducial run WA04, each with the same resolution but with a different grid size or shape.  
Figure \ref{fig:box-comp} (left panel) compares the steady state flow characteristics for various boxes with ratios $\Delta z /\Delta r \in \{1/1,2/1,4/1\}$.
While geometries and sizes with $\Delta z /\Delta r \ge 2/1$ are in excellent agreement, the quadratic domains show significantly thinner characteristics.  
The reason for this discrepancy is the sub Alfv\'enic flow that traverses the $Z_{\rm end}$ boundary in ``broad'' domains.  
In these underdetermined simulations, current circuits start to unclose at the sub Alfv\'enic part of $Z_{end}$ which ultimately destroys the Butterfly shape in the entire domain.  
As also noticed and extensively discussed by \cite{1999ApJ...526..631K}, a sub Alfv\'enic (vertical) outflow can not obtain the proper critical point information and leads to erroneous extensive collimation.
This problem can be avoided by taking the position of the critical Alfv\'en surface into account, hence we choose a ratio of $2/1$ for our science simulations.

Finally we check convergence by comparison to a half-resolution run with $256 \times 512$ grid elements.  
The solutions are in good agreement, indicated by contours of the Alfv\'en mach number in figure \ref{fig:box-comp} (right panel).  In conclusion we use a grid of $512 \times 1024$ cells with a domain size of $(r,z) = (102,204)$ inner disk radii, ensuring that the presented results depend mostly on the disk corona boundary.  

\section{The injection boundary in MHD-jet simulations}\label{sec:MHD-inj}
The number of constraints imposed on a given boundary must equal the number of 
waves allowed to travel through the boundary to the simulation domain 
(e.g. \cite{Bogovalov1997}). 
In MHD the number of characteristics equals the number of ``variables'' minus 
one - due to the $\nabla \cdot \mathbf{B}=0$ constraint - to give a total number
of seven.  

Another way of looking at this is by considering the critical surfaces as 
internal boundaries.  
For example, in a sub-sonic flow, 
the intersection of the two characteristics $C_{\pm}$ with respective wave velocities $v\pm c_{s}$ 
determines the hydrodynamical state.  
With respect to any sub-sonic boundary condition, only one characteristics is incoming, 
while the outgoing characteristics originates at the sonic point where its velocity vanishes.
The same is true for the supersonic case, 
only now $C_{-}$ transports the information of the sonic point downstream.  

Applying a $x-t$ diagram one may understand why the sonic point constitutes a fixed-in-time boundary condition.
That is because here $C_{-}$ becomes singular, and hence this part of information 
(the Riemann-invariant) starts to travel up- and down-stream from the critical 
point \citep[see also][chap.~X]{1959flme.book.....L}.  

Following these general considerations, we see that the correct number of constraints for a sub-magnetoslow boundary is four, since the flow is expected to pass through three 
characteristics. 
In other words, there are four outgoing waves: the slow magnetosonic wave, the entropy wave, 
the Alfv\'en wave and the fast magnetosonic wave.  
Naturally, along a boundary where the flow is super magneto-slow this number equals five.  
Within this limited freedom, those boundary conditions which best prescribe the 
astrophysical problem should be used.

To allow the outflow to settle into a steady state requires certain conditions to be met also at the boundary.  
In our paper, we follow the argument by \citet{1999ApJ...526..631K}.
According to the (axially symmetric) induction equation it is
\begin{equation}
	\partial_{t} B_{z} = 1/r \partial_{r} E_{\phi};\hspace{1.cm}
	\partial_{t} B_{r} = - \partial_{z} E_{\phi};\hspace{1.cm}
	\partial_{t} B_{\phi} = \partial_{z} E_{r} - \partial_{r} E_{z}.
\end{equation}
Since in steady state $\partial_{t}B_{z} = \partial_{t}B_{r}=0$, 
the only physical solution is $E_{\phi}=0$ which is satisfied by $\mathbf{v_{p}||B_{p}}$.  

A number of authors prescribe in addition a fixed-in-time value for $E_{r}=\Omega^F B_{z}$, 
however, this is equivalent to keeping the iso-rotation parameter $\Omega^F$ constant 
in time, as the time evolution of $B_{z}$ is already suppressed by the choice of $E_{\phi}$.   
Unlike often stated, the certainly proper choice of constraining $(E_{\phi},E_{r})$ is not
dictated by the ideal MHD condition - vanishing electric fields in the co-moving 
frame - (which is respected by design), 
but by the steady-state considerations given above.  
In the literature of ``disk-as-boundary'' jet formation simulations a variety of choices for the 
injection boundary exist.  
Table~\ref{tab:injections} reviews a couple of them in chronological order.

\begin{deluxetable}{lcccc}
\tablecaption{Injection conditions for disk-as boundary simulations\label{tab:injections}}
\tabletypesize{\small}
\tablewidth{\textwidth}
\tablehead{
\colhead{Authors} &
\colhead{sub/super slow} &
\colhead{Outgoing waves} &
\colhead{Constraints} &
\colhead{Nature of constraints}
}
\startdata
\cite{1995ApJ...439L..39U}\tablenotemark{a} &
sub &
4 &
4 &
$E_{\phi},E_{r},v_{z},s$
\\
\cite{1997ApJ...482..712O}\tablenotemark{b} &
sub/super &
4 &
6 &
$E_{\phi}, v_{\phi}, v_{p}, B_{\phi}, \rho, s$
\\
\cite{Romanova1997}\tablenotemark{c} &
sub &
3 &
3 &
$E_{\phi}, E_{r}, v_{z}$
\\
\cite{1999ApJ...526..631K}\tablenotemark{d} &
super &
5 &
5 &
$E_{\phi},E_{r},\rho,s,v_{z}$
\\
\cite{1999ApJ...516..221U}\tablenotemark{e} &
sub &
4 - 5 &
4 &
$E_{\phi},E_{r},\rho,s$
\\
\cite{2007MNRAS.380...51K}\tablenotemark{f} &
super &
5 &
5 &
$E_{\phi},B^\eta,\rho,v^\eta,\Omega^F$
\\
This work\tablenotemark{g} &
sub - super &
4 - 5 &
4 - 6 &
$E_{\phi},\Omega^F,\rho,s\ [, v_{z}, B_{\phi}]$
\enddata


\tablenotetext{a}{In the work of \cite{1995ApJ...439L..39U}, 
  mass flux as free parameter by allowing the disk density to evolve.}

\tablenotetext{b}{
\cite{1997ApJ...482..712O,2002A&A...395.1045F} allow no feedback from the jet to the disk and 
 seem to over-determine their simulations.  
 We were able to reproduce the consequences with our own simulations.  
 A numerical boundary layer develops which creates a steep gradient $\partial_{z}\rho(r,z)$ 
 as the code tries to match the dense disk-boundary with the jet-solution (their figure 4). 
 To conserve mass flux, the poloidal velocity will jump within a few grid cells, a spurious
 acceleration which is independent of resolution.  Due to their additional Alfv\'en pressure,
 the injection velocity is not entirely clear to us, but we suspect it to be dynamically sub-slow.  
}

\tablenotetext{c}{
\cite{Romanova1997} perform isothermal simulations without solving the energy
 equation.  
 Since the sub-slow injection, constraints on the following three quantities are given,
 $v_{z}, E_{\phi}, E_{r}$  - just as \cite{1995ApJ...439L..39U}.  
}

\tablenotetext{d}{
\cite{1999ApJ...526..631K} prescribe mass flux with the choice $(E_{\phi},E_{r},s,\rho,v_{z})$ 
 and super-slow injection. 
}

\tablenotetext{e}{
\cite{1999ApJ...516..221U} prescribe density $\rho$ instead of the vertical velocity $v_{z}$
 (thus change from \cite{1995ApJ...439L..39U}).
 The injection speed is allowed to become super-sonic which strictly speaking results in an 
 underdetermined system. 
}

\tablenotetext{f}{
\cite{2007MNRAS.380...51K} also inject super-slow.  
 As the energy equation is not solved, one condition is already used for fixing the entropy. 
 Constraints in the injection boundary *far from the disk) disk are 
 $(E_{\phi},B^\eta,\rho,v^\eta,\Omega^F)$ where the $\eta$-coordinate describes a ``radial'' 
 direction ($\eta^2 = r^2/a+z^2$, $a\ge 1$) in elliptical coordinates.
}
\tablenotetext{g}{
See the discussion in section \ref{sec:injection} of this paper.  
}

\end{deluxetable}


\clearpage


\end{document}